\documentclass[12pt,a4paper,notitlepage]{article}

\usepackage{amsfonts}
\usepackage{amsmath}
\usepackage{amssymb}
\usepackage{amsthm}
\usepackage{color}
\usepackage[T1]{fontenc}
\usepackage{graphicx}
\usepackage{hyperref}
\usepackage{mathrsfs}
\usepackage{multirow}
\usepackage[square]{natbib}
\usepackage[format=hang]{subcaption}
\usepackage{times}
\usepackage{upgreek}
\usepackage[utf8]{inputenc}
\usepackage{pgfplots}
\usepackage{tabularx}
\usepackage{xcolor}
\usepackage{pgfplotstable}
\usepackage{multicol}
\usepackage{nomencl}
\makenomenclature
\DeclareUnicodeCharacter{2212}{−}
\usepgfplotslibrary{groupplots,dateplot}
\usetikzlibrary{patterns,shapes.arrows}
\pgfplotsset{compat=newest}
\usepackage{float}
\pgfplotsset{compat=newest}
\usetikzlibrary{plotmarks}
\usetikzlibrary{arrows.meta}
\usetikzlibrary{calc}
\usepgfplotslibrary{patchplots}
\usepgfplotslibrary{external} 
\usepackage{grffile}
\AtBeginDocument{}

\newcolumntype{Y}{>{\centering\arraybackslash}X}

\definecolor{rwth1}{RGB}{0,84,159}
\definecolor{rwth2}{RGB}{246,168,0}
\definecolor{rwth3}{RGB}{0,97,101}
\definecolor{rwth4}{RGB}{204,7,30}
\definecolor{rwth5}{RGB}{87,171,39}
\definecolor{rwth6}{RGB}{161,16,53}
\definecolor{rwth7}{RGB}{0,152,161}
\definecolor{rwth8}{RGB}{122,111,172}
\pgfplotscreateplotcyclelist{rwth}{
{rwth1},{rwth2},{rwth3},{rwth4},{rwth5},{rwth6},{rwth7},{rwth8}
}

\usepackage{tikz}
\pgfrealjobname{Paper01_Review} % Im Hauptverzeichnis: 1) pdflatex Paper01_Review 2) pdflatex --jobname 'Path beginpgfgraphicnamed' Paper01_Review
\usepackage{tikz-3dplot}
\usetikzlibrary{patterns,decorations.pathmorphing}
\usepackage{environ}
\makeatletter
\newsavebox{\measure@tikzpicture}
\NewEnviron{scaletikzpicturetowidth}[1]{%
  \def\tikz@width{#1}%
  \begin{lrbox}{\measure@tikzpicture}%
  \BODY
  \end{lrbox}%
  \pgfmathparse{#1/\wd\measure@tikzpicture}%
  \BODY
}
\makeatother

\usepgfplotslibrary{fillbetween}
\pgfdeclarelayer{bg}
\pgfsetlayers{bg,main}
\usetikzlibrary{shapes.multipart}
\usetikzlibrary{spy}  

\tikzset{
    hatch distance/.store in=\hatchdistance,
    hatch distance=10pt,
    hatch thickness/.store in=\hatchthickness,
    hatch thickness=0.3pt
}

\makeatletter
\pgfdeclarepatternformonly[\hatchdistance,\hatchthickness]{northeast}
{\pgfqpoint{0pt}{0pt}}
{\pgfqpoint{\hatchdistance}{\hatchdistance}}
{\pgfpoint{\hatchdistance-1pt}{\hatchdistance-1pt}}%
{
    \pgfsetcolor{\tikz@pattern@color}
    \pgfsetlinewidth{\hatchthickness}
    \pgfpathmoveto{\pgfqpoint{0pt}{0pt}}
    \pgfpathlineto{\pgfqpoint{\hatchdistance}{\hatchdistance}}
    \pgfusepath{stroke}
}

\date{}

\graphicspath{{./images/}}

\newcounter{remark}
\newtheorem*{remark}{Remark}

\begin{document}

\author{\large {L. Lamm${}^{*,\dag}$, H. Holthusen${}^{\,\dag}$, T. Brepols${}^{\,\dag}$, S. Jockenh\"ovel${}^{\,\ddag}$, S. Reese${}^{\,\dag}$}\\[0.5cm]
  \hspace*{-0.1cm}
  \normalsize{\em ${}^{\dag}$Institute of Applied Mechanics, RWTH Aachen
    University,}\\
  \normalsize{\em Mies-van-der-Rohe-Str.\ 1, 52074 Aachen, Germany}\\[0.25cm]
  %\normalsize{\em \{tim.brepols, stefanie.reese\}@rwth-aachen.de}\\[0.25cm]
  \normalsize{\em ${}^{\ddag}$ Biohybrid \& Medical Textiles, Institute of Applied Medical Engineering, RWTH Aachen University,}\\
  \normalsize{\em Pauwelsstr.\ 20, 52074 Aachen, Germany}\\
  %\normalsize{\em swu@tf.uni-kiel.de}
}

\title{\LARGE A macroscopic approach for stress driven anisotropic growth in bioengineered soft tissues}
\maketitle

\small
{\em ${}^{*}$ Corresponding author}\\
\newline
{\bf Abstract.}
The simulation of growth processes within soft biological tissues is of utmost importance for many applications in the
medical sector. Within this contribution we propose a new macroscopic approach fro modelling stress-driven volumetric growth
occurring in soft tissues. Instead of using the standard approach of a-priori defining the structure of the growth tensor,
we postulate the existance of a general growth potential. Such a potential describes all eligable homeostatic stress states
that can ultimately be reached as a result of the growth process. Making use of well established methods from
visco-plasticity, the evolution of the growth related right Cauchy-Green tensor is subsequently defined as a time
dependent associative evolution law with respect to the introduced potential. This approach naturally leads
to a formulation that is able to cover both, isotropic and anisotropic growth related changes in geometry. It
furthermore allows the model to flexibly adapt to changing boundary and loading conditions. Besides the theoretical
development, we also describe the algorithmic implementation and furthermore compare the newly derived model with a
standard formulation of isotropic growth.

\vspace*{0.3cm}

\vspace*{0.3cm}
{\bf Acknowledgements:} {L. Lamm, J. Jockenv\"ovel and S. Reese gratefully acknowledge the financial
  support provided by the German Research Foundation (DFG) within the subproject ‘Modelling of the structure
  and fluid–structure interaction of biohybrid heart valves on tissue maturation’ of the DFG PAK 961
  ‘Towards a model based control of biohybrid implant maturation’ (RE 1057/45-1, Project number 403471716). Furthermore, H. Holthusen
  and S. Reese acknowledge the financial support through the DFG Project 'Experimental and numerical
  investigation of layered, fiber-reinforced plastics under crash loads' (RE 1057/46-1, Project number 404502442). L. Lamm, T. Brepols and
  S. Reese acknowledge the support granted by the AiF through the Project 'Methodology for dimensioning and
  simulation of adhesive bonds in glass-facade structures' (IGF 21348 N/3).}

\normalsize

%%%%%%%%%%%%%%%%%%%%%%%%%%%%%%%%%%%%%%%%%%%%%%%%%%%%%%%%%%%%%%%%%%%%%%%%%%%%%%%%%%%%%%%%%%%%%%%%%%%%%%%%%

\section{Introduction}
\label{sec:1}

The production and use of artificially grown biological tissue has become an important research
topic in the medical context over the last two decades. Great progress has been made in implant
research in particular, with the cultivation of biohybrid heart valves being just one example
among many (\cite{Fioretta_2019}). Designing and constructing highly complex medical implants
is a big challenge due to the biomechanical properties of the underlying cultivated tissue.
Early works in the field of biomechanics have already pointed out that biological tissues adapt
dynamically to the environment they are exposed to (see e.g. \cite{Fung_1995} and references therein).
The goal of this process is to reach a homeostatic state in which e.g. a certain critical stress state
is neither exceeded nor fallen below. During this process, which we will call growth in the following,
the progression towards a homeostatic stress state is mainly driven by a change in mass and internal
structure of the given biological material. Subsequently, such a growth process leads to a change in
the mechanical behaviour, which usually has a large influence on the performance of the given implant.
In contrast to native tissue, these adaptive effects are particularly pronounced during the
cultivation period of bioengineered tissues and must therefore be taken into account from the
beginning of the design process. Within this context, computational modelling contributes to a
deeper understanding and prediction of such adaptation processes. An important aspect of modelling
the mechanics of growth is the description of geometry changes which are due to contraction and
expansion of the material, respectively. Starting from the works of \cite{Skalak_1981}, \cite{Skalak_1982}
and \cite{Rodriguez_EtAl_1994}, many models have been developed over the last decades in order to describe
such finite volumetric growth effects. Although being already successfully applied e.g. in the
modelling of finite plasticity (see e.g. \cite{Eckart_1948}, \cite{Kroener_1959}, \cite{Lee_1969}),
it was the contribution of \cite{Rodriguez_EtAl_1994} which first adapted the multiplicative
split of the deformation gradient to describe the inelastic nature of finite growth processes.
For a detailed overview on the various modelling strategies, the interested reader is referred
to the comprehensive overviews given e.g. by  \cite{Goriely_2017} and \cite{Ambrosi_2019}.
Most of the approaches based on the conceptually simple and computationally efficient framework by
\cite{Rodriguez_EtAl_1994} can be roughly divided into two different groups, isotropic (e.g.
\cite{Lubara_Hoger_2002}, \cite{Himpel_EtAl_2005}) and anisotropic growth models (e.g. \cite{Menzel_2005},
\cite{Goektepe_EtAl_2010}, \cite{Soleimani_2020}). It is important to outline that the distinction
between isotropic and anisotropic growth is by no means related to the underlying elastic material
behaviour, but rather describes the way a referential volume element changes its shape over time.
In case of isotropic growth, the deformation gradient tensor is often assumed to be proportional to
the identity tensor (e.g. \cite{Lubara_Hoger_2002}), which yields a uniform expansion of a referential
volume element. On the other hand, the term anisotropic growth describes a geometry change of a given
volume element that is not uniform in all three spatial dimensions but rather has a distinct growth
direction (e.g. \cite{Goektepe_EtAl_2010}). Despite its widespread use, the approach of isotropic
growth modelling has strong limitations with regard to describing the mechanical behaviour of
complex structures. Recently, \cite{Braeu_2017} and \cite{Braeu_2019} pointed out that in the context
of relevant applications, anisotropic growth behaviour is more the standard case than an isotropic
response. Classically, this intrinsically anistropic growth behaviour is modelled using heuristic
assumptions on the definition of preferred growth directions. This, unfortunately, yields the need to
a-priori prescribe a certain structure of the growth related deformation gradient. Whilst this approach
might be feasible for relatively simple problems such as e.g. fibre elongation and contraction,
it is not well applicable for more complex applications. It is therefore still an ongoing topic of
research to define a more general and flexible formulation that is able to adapt to various boundary
value problems, as pointed already out by e.g. \cite{Menzel_2005} or \cite{Soleimani_2020}.
In addition to the phenomenologically motivated models described above, another class of models was
established for describing growth processes. Originating from the theory of mixtures, \cite{Humphrey_2002},
among others, developed the constrained mixture theory. Instead of assuming that the volume as a
whole is deformed during the growth process, this modelling approach describes the change of
volume in terms of a continuous deposition and removal of mass increments. Since this approach
is computationally very expensive, \cite{Cyron_2016} and \cite{Cyron_2017} developed a
homogenized version of the constrained mixture model. This is achieved by using a temporal
homogenization of the mass increments alongside with the same multiplicative split as described
by \cite{Rodriguez_EtAl_1994}. Although this approach overcomes the limitations of the classical
constrained mixture theory in terms of computational costs, it still suffers from the need to
a-priori define the structure of the growth tensor.

As pointed out in the brief literature review above, most existing models in the field of growth
simulations require to make assumptions on the structure of the growth tensor.
This leads to a situation where growth can always only evolve in a predefined manner. Even if the
underlying growth process is the same, this can possibly result in a scenario where the same material
subjected to different loading or boundary conditions might need a different constitutive model in
order to capture the macroscopical behaviour properly. Consider, e.g., a piece of engineered soft
collagenous tissue. It is well known that such a material tends to shrink during its maturation phase,
in order to reach homeostasis. The way in which this shrinking process takes place depends very much
on the given boundary conditions. While a specimen that is not hindered in its deformation by external
boundary conditions clearly exhibits isotropic deformation behaviour, this is no longer the case as soon
as, e.g., two sides of the specimen are clamped (see e.g. experiments in \cite{Gauvin_2013} and
\cite{Ghazanfari_2015}). In the authors' opinion, a constitutive formulation should be developed which
is capable of simulating growth processes independently of the given boundary conditions. The present work
therefore introduces a novel and flexible framework for the description of stress driven volumetric growth.
This framework does not depend on the a-priori definition of the growth tensors structure and is able to
cover both, isotropic and anisotropic growth behaviour, naturally. Section~\ref{sec:2} covers the
theoretical modelling ideas behind the proposed model. The numerical implementation of the derived
material model is described in Section~\ref{sec:3}. Finally, numerical examples are given in
Section~\ref{sec:4}.

\nomenclature[1a]{$a$}{Scalar}
\nomenclature[1b]{$\mathbf{a}$}{First order tensor or vector}
\nomenclature[1c]{$\mathbf{A}$}{Second order tensor or matrix}
\nomenclature[1f]{$\mathbb{A}$}{Fourth order tensor}
\nomenclature[80a]{$(\bullet)_e$}{Elastic quantity}
\nomenclature[80b]{$(\bullet)_g$}{Growth related quantity}
\nomenclature[99a]{$\cdot$}{Single contraction ($=:\mathbf{A}\mathbf{B}$)}
\nomenclature[99b]{$:$}{Double contraction}
\nomenclature[99c]{$\otimes$}{Dyadic product}
\nomenclature[99cc]{$\mathbf{A}^T$}{Transpose of $\mathbf{A}$}
\nomenclature[99cd]{$\mathbf{A}^{-1}$}{Inverse of $\mathbf{A}$}
\nomenclature[99dd]{$\operatorname{tr}{\mathbf{A}}$}{Trace of $\mathbf{A}$}
\nomenclature[99de]{$\operatorname{det}\mathbf{A}$}{Determinant of $\mathbf{A}$}
\nomenclature[99df]{$\operatorname{dev}{\mathbf{A}}$}{Deviator of $\mathbf{A}$}
\nomenclature[99dg]{$\operatorname{sym}{\mathbf{A}}$}{Symmetric part of $\mathbf{A}$}
\nomenclature[99dh]{$\operatorname{exp}(\bullet)$}{Exponential function}
\nomenclature[99e]{$||(\bullet)||$}{Frobenius norm}
\nomenclature[99i]{$\operatorname{Grad}(\bullet)$}{Lagrangian gradient}
\nomenclature[99j]{$\operatorname{Div}(\bullet)$}{Divergence}
\nomenclature[99k]{$\frac{\partial (\bullet)}{\partial (\bullet)}$}{Partial derivative}
\nomenclature[99kk]{$\Delta(\bullet)$}{Increment of $(\bullet)$}
\nomenclature[99l]{$\dot{(\bullet)}$}{Total time derivative}
\nomenclature[99zzzz]{$\hat{(\bullet)}$}{Voigt notation $(11,22,33,12,13,23)$}

\begin{multicols}{2}
    \printnomenclature[0.7in]
\end{multicols}
%makeindex main.nlo -s nomencl.ist -o main.nls

%%% Local Variables: 
%%% mode: latex
%%% TeX-master: "../main"
%%% End: 

\section{Continuum mechanical modelling of finite growth}
\label{sec:2}

Let us first introduce the well established multiplicative split of the deformation gradient \(\mathbf{F}\)
into an elastic and a growth related part (see e.g. \cite{Rodriguez_EtAl_1994}), i.e.
\begin{equation}
    \label{eq:1}
    \mathbf{F} = \mathbf{F}_e\mathbf{F}_g.
\end{equation}
Using this equation, the determinant of \(\mathbf{F}\), abbreviated by \(J:=\operatorname{det} \mathbf{F} = \operatorname{det} \mathbf{F}_e \operatorname{det} \mathbf{F}_g\),
is also multiplicatively split into two parts. Whilst the change of volume due to elastic deformations
is described by \(J_e = \operatorname{det} \mathbf{F}_e\), the growth related volume changes are represented
by \(J_g = \operatorname{det} \mathbf{F}_g\). In analogy to the right Cauchy-Green tensor
\(\mathbf{C} = \mathbf{F}^T\mathbf{F}\) as well as the left Cauchy-Green tensor \(\mathbf{B}=\mathbf{F}\mathbf{F}^T\),
the elastic right Cauchy-Green tensor and the growth related right and left Cauchy-Green tensor can be
defined as
\begin{equation}
    \label{eq:2}
    \begin{split}
        \mathbf{C}_e &:=\mathbf{F}_e^T\mathbf{F}_e = \mathbf{F}_g^{-T}\mathbf{C}\mathbf{F}_g^{-1}\\
        \mathbf{C}_g &:=\mathbf{F}_g^T\mathbf{F}_g\\
        \mathbf{B}_g &:=\mathbf{F}_g\mathbf{F}_g^T.
    \end{split}
\end{equation}
Furthermore, the growth related velocity gradient \(\mathbf{L}_g\) is introduced as
\begin{equation}
    \label{eq:3}
    \mathbf{L}_g = \dot{\mathbf{F}}_g\mathbf{F}_g^{-1}.
\end{equation}

%------------------------------------------------------------------------------------------------------------------%
\subsection{Balance relations}
\label{sec:2-1}
Growth processes within biological systems in general lead to a change of the systems mass as well as a
change of its shape and volume, respectively. Within this contribution, the focus lies on the macroscopic
description of changes in shape rather than a change of the systems mass. We therefore neglect the
description of the balance of mass in terms of production or flux terms and assume that this balance
relation is fulfilled implicitly. It is furthermore  well established to assume that growth processes
take place on a significantly larger time scale than mechanical deformations do. This standard argument
is known as the \textit{slow growth assumption} and yields for a quasi-static setup of the well known
balance of linear momentum
\begin{equation}
    \label{eq:4}
    \operatorname{Div}\left(\mathbf{F}\mathbf{S}\right) + \mathbf{b}_0 = \mathbf{0}.
\end{equation}
Here \(\mathbf{S}\)  and \(\mathbf{b}_0\) denote the second Piola-Kirchhoff stress tensor and
the referential body force vector per reference volume, respectively. Following the idea of open system
thermodynamics (see e.g. \cite{Kuhl_Steinmann_2003b} and references therein), we describe the entropy
production \(\dot{\gamma}\) in terms of the Clausius-Duhem inequality
\begin{equation}
    \label{eq:5}
    \dot{\gamma} = \mathbf{S} : \frac{1}{2}\dot{\mathbf{C}} - \dot{\psi} +  \mathcal{S}_0 \geq 0,
\end{equation}
with the volume specific Helmholtz free energy density \(\psi\) defined more precisely in the following
section. It is important to notice, that the additional referential entropy sink \(\mathcal{S}_0\) is
introduced to allow for a decrease in entropy due to the growth process itself. Therefore, \(\mathcal{S}_0\)
is never actually computed.

%------------------------------------------------------------------------------------------------------------------%
\subsection{Helmholtz free energy}
\label{sec:2-2}
We start from the general continuum mechanical framework laid down in \cite{Svendsen_2001}. Within this
context, the constitutive equations are described with respect to a given but otherwise arbitrary
configuration of the material body in question. Similar to the approaches made by \cite{Bertram_1999}
and \cite{Svendsen_2001} in the context of finite plasticity, we choose the elastic part of the Helmholtz
free energy to be stated in terms of quantities defined within the so-called grown intermediate configuration.

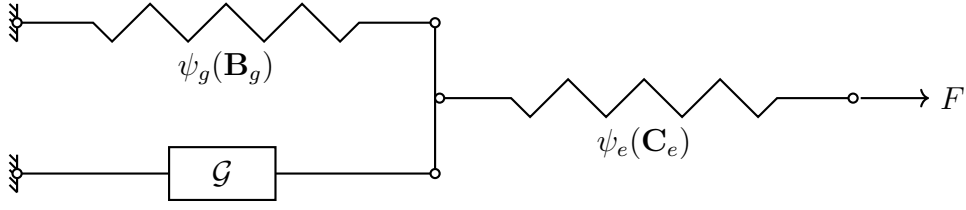
\begin{figure}
    \centering
    \begin{tikzpicture}
    % Draw Backstress Spring Element and right line
    \draw [thick] (0,2) -- (1,2) -- (1.25,1.75) -- (1.75,2.25) -- (2.25,1.75) -- (2.75,2.25) --(3.25,1.75) -- (3.75,2.25) --(4.25,1.75) -- (4.5,2) --(5.5,2);
    \draw [thick] (5.5,2) -- (5.5,0);
    \draw [thick] (0,0) -- (5.5,0);
    \draw [thick, fill=white, radius=0.06] (5.5,2) circle;
    \draw [thick, fill=white, radius=0.06] (5.5,0) circle;

    % Draw growth element
    \filldraw [fill=white, thick] (2.0, -0.35) rectangle (3.4, 0.35);
    \node at (2.7,0) {\(\mathcal{G}\)};

    % Draw elastic spring
    \draw [thick] (5.5,1) -- (6.5,1) -- (6.75,0.75) -- (7.25,1.25) -- (7.75,0.75) -- (8.25,1.25) --(8.75,0.75) -- (9.25,1.25) --(9.75,0.75) -- (10,1) -- (11,1);
    \draw [thick, fill=white, radius=0.06] (11,1) circle;
    \draw [thick, fill=white, radius=0.06] (5.56,1) circle;

    % supports on the left side
    \foreach \y in {-0.25, 1.75}
        {
            % Define corner point A for support
            \coordinate (A) at (0,\y);
            % Draw vertical line
            \draw [thick] (A) -- ++(0,0.5);
            % Draw diagonal lines
            \foreach \yy in {0, 0.1, 0.2, 0.3, 0.4}
                {
                    \draw [thick] ($ (A) + (0, \yy) $) -- ++(-0.1,0.1);
                }
            % Draw circle
            \draw [fill=white, thick] ($ (A) + (0,0.25) $) circle [radius=0.06];;
        }

    % Draw loading 
    \draw [thick, ->] (11.1,1) -- (12,1);

    % Draw annotations
    \node[right] at (12,1) {\(F\)};
    \node[] at (2.75,1.4) {\(\psi_g(\mathbf{B}_g)\)};
    \node[] at (8.25,0.4) {\(\psi_e(\mathbf{C}_e)\)};

\end{tikzpicture}
    \caption{Rheological model corresponding to the given volumetric growth model. Growth is denoted by
        the element including the character \(\mathcal{G}\).}
    \label{fig:2-99}
\end{figure}

When modelling finite volumetric growth, it is important to ensure that infinite expansion or shrinkage is avoided.
A common approach to tackle this problem is to introduce a set of limiting material parameters,
which can be interpreted as the maximum possible growth induced stretches (see e.g. \cite{Lubara_Hoger_2002}).
Although such approaches may give computationally reasonable results, this assumption cannot be motivated by
underlying physical phenomena.
Within the context of growth processes taking place in engineered biological tissues, it is reasonable to
assume that a growth related change in volume is always accompanied by a rise in internal pressure.
Such pressure accumulations are consequently counteracting the expansion and contraction process, respectively. This growth related internal pressure
can be described by including an additional dependency on either \(\mathbf{C}_g\) or \(\mathbf{B}_g\).
Using the idea of interpreting \(\mathbf{F}_g\) as a so-called \textit{material isomorphism} (see e.g.
\cite{Noll_1958}, \cite{Svendsen_2001}), it follows that one has to choose \(\mathbf{B}_g\) in order to
ensure that the kinematic quantities are located within the same configuration, i.e.
\begin{equation*}
    \psi := \tilde{\psi}\left(\mathbf{C}_e, \mathbf{B}_g\right).
\end{equation*}
Note that this choice is strongly related to the general concept of structural tensors. In the present case,
namely by choosing the structural tensor equal to \(\mathbf{B}_g\), the relation to linear kinematic
hardening becomes obvious. This is worked out in the paper of \cite{Dettmer_Reese_2004}, where linear
kinematic hardening is a special case of the so-called Armstrong-Frederick type of kinematic hardening.
In the following, we choose an additive format, i.e.
\begin{equation}
    \label{eq:6}
    \tilde{\psi} := \psi_e(\mathbf{C}_e) + \psi_g(\mathbf{B}_g),
\end{equation}
for the Helmholtz free energy, where the the both terms are isotropic functions of \(\mathbf{C}_e\) and
\(\mathbf{B}_g\), respectively. Such an additive split can be motivated easily by the rheological model
shown in Figure~\ref{fig:2-99}.
This model illustrates nicely that a growth related expansion or contraction directly
results in an accumulation of the growth related energy \(\psi_g\) due to the loading of the
associated spring element. Such an increase in growth related energy clearly counteracts the
growth deformation and ultimately leads to a decaying growth response.
The elastically stored energy \(\psi_e\) is represented within this rheological
model by the second spring element. It is obvious that this particular spring is influenced by both,
growth related and purely elastic deformations. Therefore, any deformation associated with the growth
element will lead to a change in the elastically stored energy, even under the absence of external loadings.

%------------------------------------------------------------------------------------------------------------------%
\subsection{Thermodynamic considerations}
\label{sec:2-3}
To derive the constitutive equations representing finite volumetric growth, we next consider the isothermal
Clausius-Duhem inequality as given in Equation~\eqref{eq:5}. Inserting the Helmholtz free energy
(Equation~\eqref{eq:6}) and differentiating with respect to time yields
\begin{equation}
    \label{eq:7}
    \mathbf{S} : \frac{1}{2}\dot{\mathbf{C}} - \left(\frac{\partial\psi_e}{\partial\mathbf{C}_e} : \dot{\mathbf{C}}_e +
    \frac{\partial\psi_g}{\partial\mathbf{B}_g}:\dot{\mathbf{B}}_g\right) +  \mathcal{S}_0 \geq 0.
\end{equation}
By using the product rule as well as utilizing the identities \(\dot{\overline{\mathbf{F}_g^{-T}}} = -\mathbf{F}_g^{-T}\dot{\overline{\mathbf{F}_g^{T}}}\mathbf{F}_g^{-T}\)
and \(\dot{\overline{\mathbf{F}_g^{-1}}} = -\mathbf{F}_g^{-1}\dot{\mathbf{F}}_g\mathbf{F}_g^{-1}\),
the elastic deformation rate can be expressed as
\begin{equation}
    \label{eq:8}
    \dot{\mathbf{C}}_e = \mathbf{F}_g^{-T}\dot{\mathbf{C}}\mathbf{F}_g^{-1}
    - \mathbf{L}_g^T\mathbf{F}_g^{-T}\dot{\mathbf{C}}\mathbf{F}_g^{-1} -
    \mathbf{F}_g^{-T}\dot{\mathbf{C}}\mathbf{F}_g^{-1}\mathbf{L}_g.
\end{equation}
With the definition of the growth velocity gradient given in Equation~\eqref{eq:3}, the growth related
deformation rate can similarly be found as
\begin{equation}
    \label{eq:9}
    \dot{\mathbf{B}}_g = \mathbf{L}_g\mathbf{B}_g + \mathbf{B}_g\mathbf{L}_g^T.
\end{equation}
Combining the equations above and making use of the standard procedure of \cite{Coleman_Noll_1963},
we can find the thermodynamically consistent definition of the second Piola-Kirchhoff stress tensor as
\begin{equation}
    \label{eq:10}
    \mathbf{S} = 2\mathbf{F}_g^{-1}\frac{\partial\psi}{\partial\mathbf{C}_e}\mathbf{F}_g^{-T}.
\end{equation}
Next we introduce
the Mandel stress tensor \(\mathbf{M} = 2\mathbf{C}_e\frac{\partial\psi}{\partial\mathbf{C}_e}\) and
the back-stress tensor \(\boldsymbol{\chi} = 2\mathbf{B}_g\frac{\partial\psi_2}{\partial\mathbf{B}_g}\).
Since the free energy functions \(\psi_e\) and \(\psi_g\) are chosen as
isotropic function of \(\mathbf{C}_e\) and \(\mathbf{B}_g\), respectively, their derivatives
\(\frac{\partial\psi_e}{\partial\mathbf{C}_e}\) and \(\frac{\partial\psi_g}{\partial\mathbf{B}_g}\) are
symmetric and commute with either \(\mathbf{C}_e\) or \(\mathbf{B}_g\). Combining this with the properties of the
double contracting product, the reduced Clausius-Duhem inequality can be written only in terms of the
symmetric part \(\mathbf{D}_g = \operatorname{sym}\mathbf{L}_g\) of the growth velocity gradient, i.e.
\begin{equation}
    \label{eq:11}
    \dot{\gamma}_{red} = \left[\mathbf{M} - \boldsymbol{\chi}\right] : \mathbf{D}_g + \mathcal{S}_0 \geq 0.
\end{equation}
Similar to classical plasticity theory, see e.g. \cite{Vladimirov_EtAl_2008}, one can identify the Mandel
stress tensor \(\mathbf{M}\) and the back-stress tensor \(\boldsymbol{\chi}\) as the conjugated driving
forces for the evolution of growth. It is therefore natural to describe the evolution equation for
\(\mathbf{D}_g\) in terms of these quantities. Notice that \(\mathbf{M}\) and \(\boldsymbol{\chi}\) are
located within a grown intermediate configuration, where they can be clearly identified as stress like
quantities. This becomes clear by the fact that \(\mathbf{M}\) has the same invariants as the Kirchhoff
stress tensor \(\boldsymbol{\tau}\) and, thus, has a clear physical meaning (see Appendix~\ref{sec:Appendix-1}).
Pulling \(\mathbf{M}\) and \(\boldsymbol{\chi}\) back to the reference configuration will yield a loss of such
clear physical interpretation. Nevertheless, from a conceptual and computational point of view, a pull
back of of these quantities to the reference configuration is desirable (for details see e.g.
\cite{Dettmer_Reese_2004} and \cite{Vladimirov_EtAl_2008}). Taking into account the relation
\(\mathbf{D}_g = \frac{1}{2}\mathbf{F}_g^{-T}\dot{\mathbf{C}}_g\mathbf{F}_g^{-1}\) one can rewrite the
Clausius Duhem inequality purely in terms of quantities located within the reference configuration, i.e.
\begin{equation}
    \label{eq:12}
    \begin{split}
        \dot{\gamma}_{red} &= \left(\mathbf{F}_g^{-1}\mathbf{M}\mathbf{F}_g^{-T} - \mathbf{F}_g^{-1}\boldsymbol{\chi}\mathbf{F}_g^{-T}\right) : \frac{1}{2}\dot{\mathbf{C}}_g + \mathcal{S}_0\\
        &= \left(\mathbf{\Gamma} - \mathbf{X}\right) : \frac{1}{2}\dot{\mathbf{C}}_g + \mathcal{S}_0\\
        &= \mathbf{\Sigma}: \frac{1}{2}\dot{\mathbf{C}}_g + \mathcal{S}_0 \geq 0.
    \end{split}
\end{equation}
Similar to the formulation given with respect to the grown intermediate configuration, it is reasonable
to define the evolution of the growth related right Cauchy-Green tensor \(\mathbf{C}_g\) in terms of the
thermodynamically conjugated driving forces \(\mathbf{\Gamma} = \mathbf{C}_g^{-1}\mathbf{C}\mathbf{S}\) and
\(\mathbf{X}=2\frac{\partial \psi_g}{\partial\mathbf{C}_g}\). It is important to mention that using
\(\mathbf{C}_g\) as the internal variable yields the fact that \(\mathbf{F}_g\) must never be computed
in the first place. Such an approach is in clear contrast to the standard formulations in volumetric
growth modelling, where the the growth tensor itself is usually explicitly prescribed.

%------------------------------------------------------------------------------------------------------------------%
\subsection{Evolution of growth}
\label{sec:2-4}
Up to this point, the framework presented is very general and could be used to describe a wide variety of
inelastic phenomena in finite deformations. It is therefore the choice of evolution equations for
\(\mathbf{C}_g\) that explicitly defines a particular kind of inelastic material model. For the most
simple modelling assumption of a purely isotropic growth response, the inelastic part of the deformation
gradient is usually defined as \(\mathbf{F}_g = \vartheta\mathbf{I}\), where \(\vartheta\) describes the
growth induced stretch (see e.g. \cite{Lubara_Hoger_2002}, \cite{Himpel_EtAl_2005}, \cite{Goektepe_EtAl_2010}).
Using the thermodynamic framework described above, this assumption naturally leads to an evolution equation
of \(\mathbf{C}_g\), which can be written as
\begin{equation}
    \label{eq:13}
    \dot{\mathbf{C}}_g := 2\frac{\dot{\vartheta}}{\vartheta}\mathbf{C}_g.
\end{equation}
Within this context, a scalar valued evolution equation \(\dot{\vartheta} = f(\vartheta, \mathbf{\Sigma}, ...)\)
is used to determine the overall growth response (see Appendix~\ref{sec:Appendix-3} for a more detailed
example). Although the a priori assumption of \(\mathbf{F}_g\) being a diagonal tensor is tempting due
to its computational simplicity, it was already pointed out in various publications that such an assumption
is not reasonable for many applications (see e.g. \cite{Soleimani_2020}, \cite{Braeu_2019}, \cite{Braeu_2017}).
This is especially the case for scenarios in which the body cannot grow freely but is restricted by complex
boundary conditions. To overcome this issue, a new volumetric growth model is proposed in the following.

\subsubsection{Finite growth using a growth potential}
\label{sec:2-4-1}
As described in the introduction, cell mediated expansion or compaction of engineered tissues takes place
in such a way that a preferred homeostatic stress state is reached within the material. In the present work,
it is assumed that this homeostatic state can be described in terms of a scalar equivalent stress. Thus,
growth always takes place, if this equivalent stress is not equal to the preferred stress state of the
biological material. These considerations lead us to the introduction of a general growth potential
\begin{equation}
    \label{eq:14}
    \Phi := \tilde{\Phi}\left(\mathbf{M}, \boldsymbol{\chi}, \alpha_1, ..., \alpha_n\right),
\end{equation}
which is a function of the conjugated driving forces as well as a set of material parameters \(\alpha_i\).
Similar to the representation used in classical plasticity theory, this potential can be represented as a
surface, located within the principal stress space, which contains all eligible homeostatic stress states.
It will therefore be named \textit{homeostatic surface} in the following. An example for such a homeostatic
surface can be found in Figure~\ref{fig:2-1}. The overall goal of this process is to approach \(\Phi = 0\)
over time and therefore reach a stress state that lies on the homeostatic surface. Furthermore, it seems
natural that such growth processes always try to minimize the amount of energy needed to reach the
homeostatic state. Hence, the direction of growth response will be described by as the derivative of the
growth potential, i.e. \(\mathbf{N} = \frac{\partial\Phi}{\partial\mathbf{M}}\).
It is furthermore obvious that homeostasis is never reached instantaneously but rather approached over a
certain period of time. To account for this temporal effect, we introduce the growth multiplier
\(\dot{\lambda}_g := \dot{\lambda}_g\left(\Phi, \eta, \beta_1, ..., \beta_n\right)\) defined as an explicit
function of the growth potential, the growth velocity \(\eta\) and a set of material parameters \(\beta_i\).
Subsequently, the considerations above lead us to an associative growth evolution law that is postulated as
\begin{equation}
    \label{eq:15}
    \mathbf{D}_g := \dot{\lambda}_g \frac{\mathbf{N}}{||\mathbf{N}||}.
\end{equation}
In general, we do not want to restrict the choice of \(\Phi\) to only positive homogeneous potentials of
degree one. This has the side effect that \(||\mathbf{N}|| = 1\) can not be guaranteed, which yields the
need to normalize the growth direction tensor to assure that only \(\dot{\lambda}_g\) has an influence on
the amount of accumulated growth deformations. As before, we furthermore can define the given evolution
equation in terms of quantities located purely within the reference configuration. To achieve this, a
pull back operation is performed that yields
\begin{equation}
    \label{eq:16}
    \dot{\mathbf{C}}_g = \frac{2\dot{\lambda}_g}{||\mathbf{N}||}\mathbf{F}_g^T\mathbf{N}\mathbf{F}_g
    = \dot{\lambda}_g\mathbf{f}
    = \dot{\lambda}_g\mathbf{g}\mathbf{C}_g,
\end{equation}
including the general second order tensors \(\mathbf{f}=\frac{2}{||\mathbf{N}||}\mathbf{F}_g^T\mathbf{N}\mathbf{F}_g\)
as well as \(\mathbf{g} = \mathbf{f}\mathbf{C}_g^{-1}\).

\begin{remark}
    The same result for Equation~\eqref{eq:16} could also be obtained following \cite{Reese_EtAl_2021} and
    the procedures proposed therein. Therefore, this evolution law could be interpreted in the broader
    context of a theory describing the evolution of general structural tensors.
\end{remark}

Since this approach is very similar to the classical models of visco-plasticity, the attentive reader may
ask how far these approaches differ. In the case of plasticity, the yield criterion is used to clearly
distinguishe between the purely elastic and elasto-plastic state, i.e. the yield criterion must always be
less than or equal to zero. In contrast, the growth potential \(\Phi\) does not serve to distinguish between
an elastic and inelastic region, since an ’elasto-growth’ state is present for both \(\Phi < 0\) and
\(\Phi > 0\). Only in case of \(\Phi = 0\) no further growth has to take place, since homeostasis has
already been reached. This behaviour is also reflected by the growth multiplier, which in contrast to
plasticity can also have negative values. In the authors opinion this modelling approach has several
advantages: (i) As stated earlier, the direction of growth does not have to be a priori prescribed, (ii)
the complexity of the material model is reduced and (iii) due to the strong similarities to plasticity,
one can rely on a large repertoire of knowledge from this field, both from a modeling and numerical point
of view. For instance, one could argue that the preferred stress can not only be described by only one
smooth growth potential. Having e.g. the concept of multisurface plasticity in mind, it
would be easy to adopt the growth potential by a more sophisticated approach. In addition, it is also
possible, for instance, to take into account a changing preferred stress using an approach similar to the
concept of isotropic hardening.

Before defining a specific form of the growth potential, we first take a closer look at the structure of
such a potential. It has already been pointed out above that it is reasonable to assume that growth
in biological tissues tends to be of isotropic nature only in the absence of restricting boundary
conditions. This idea leads us to the definition of the growth potential as a function of the volumetric
invariant \(I_1 := \operatorname{tr}\left(\mathbf{M}-\boldsymbol{\chi}\right) = \operatorname{tr}\left(\mathbf{\Sigma}\mathbf{C}_g\right)\).
To allow also for an anisotropic growth response, we furthermore include the deviatoric invariant
\(J_2 := \frac{1}{2}\operatorname{tr}\left(\operatorname{dev}\left(\mathbf{M}-\boldsymbol{\chi}\right)^2\right)
= \frac{1}{2}\operatorname{tr}\left(\operatorname{dev}\left(\mathbf{\Sigma}\mathbf{C}_g\right)^2\right)\)
(see Appendix~\ref{sec:Appendix-2}). With theses considerations at hand, we propose a general form for the
growth potential as
\begin{equation}
    \label{eq:17}
    \Phi := \Phi\left(I_1, J_2, \omega_{hom}\right) = \phi_1\left(I_1\right) + \phi_2\left(J_2\right) - \omega_{hom}.
\end{equation}
Here, the material parameter \(\omega_{hom}\) describes a stress like quantity defining the state of homeostasis.
It is important to emphasize that the combination of \(I_1\) and \(J_2\) is crucial for the proposed
material model. If the potential was merely defined in terms of the volumetric invariant \(I_1\), the
growth direction tensor would become proportional to the identity tensor which consequently yields an
evolution equation that is similar to the isotropic evolution law given in Equation~\eqref{eq:13}. It is the
dependency on \(J_2\) that introduces an anisotropic growth behaviour, since the growth direction tensor no
longer necessarily has to correspond to the identity. Nevertheless, in case of purely volumetric stress
states, the dependency on \(I_1\) ensures the desired isotropic growth response. This consideration yields
an exclusion of any purely deviatoric potential, e.g. of von Mises type potentials. Furthermore, any suitable
potential must fulfill \(\frac{\partial\Phi}{\partial\mathbf{M}} \neq \mathbf{0}\) for all
\(\left(\mathbf{M}-\boldsymbol{\chi}\right)\in\left(\mathbb{R}^3\times\mathbb{R}^3\right)\) in order to
guarantee a well-defined growth direction for any arbitrary loading condition.

\subsubsection{Choice of the growth potential and growth multiplier evolution}
\label{sec:2-4-2}
\begin{figure}
    \centering
    \begin{subfigure}{0.75\textwidth}
        \centering
        \include{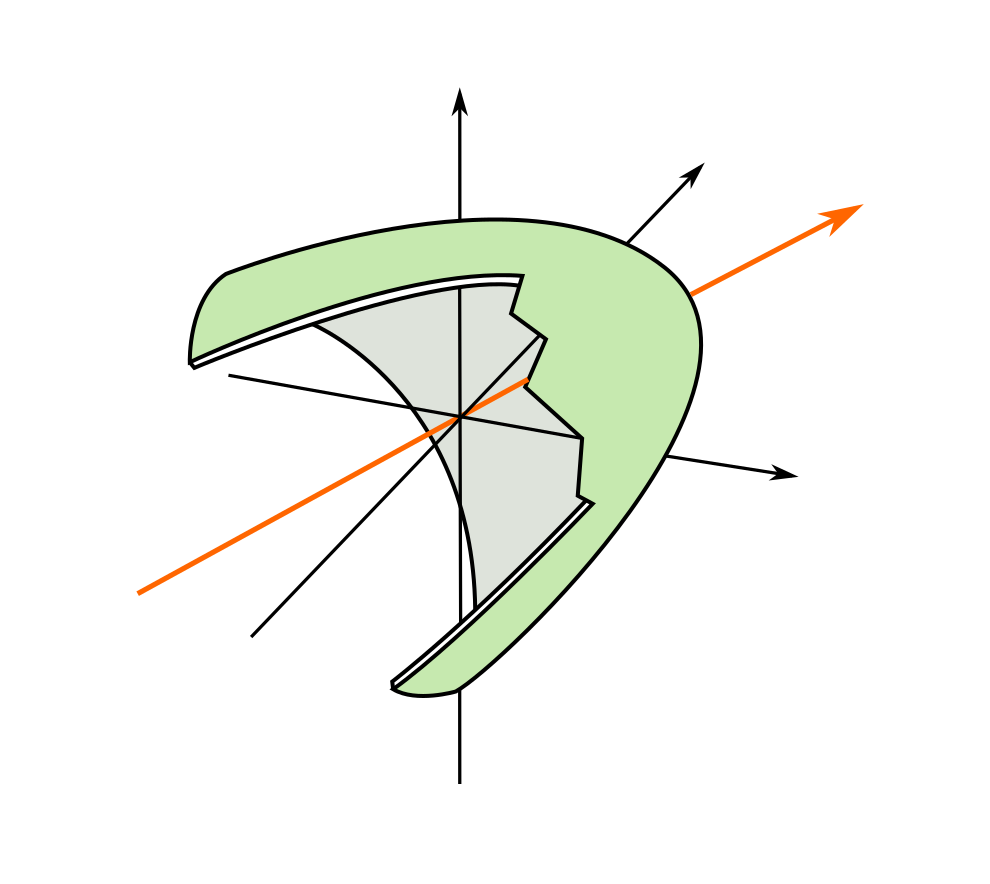}
        \caption{Homeostatic surface located in principal stress space.}
        \label{fig:2-1-a}
    \end{subfigure}
    \begin{subfigure}{0.49\textwidth}
        \centering
        \include{images/Potential/potential_smaller_one}
        \caption{Intersection of homeostatic surface with \(\sigma_1 = \sigma_2\), showing the influence of the material
            parameter \(m\) for \(m_i<1\) with \(m_1<m_2<m_3\).}
        \label{fig:2-1-b}
    \end{subfigure}
    \hfill
    \begin{subfigure}{0.49\textwidth}
        \centering
        \include{images/Potential/potential_greater_one}
        \caption{Intersection of homeostatic surface with \(\sigma_1 = \sigma_2\), showing the influence of the material
            parameter \(m\) for \(m_i>1\) with \(m_1<m_2<m_3\).}
        \label{fig:2-1-c}
    \end{subfigure}
    \caption{Schematic representation of the homeostatic surface defined by the growth potential from Equation~\eqref{eq:18}
        displayed in principal stress space. The hydrostatic axis \(p = \operatorname{tr}\left(\mathbf{M}- \boldsymbol{\chi}\right)\)
        is shown in orange. The eigenvaules of \(\mathbf{M} - \boldsymbol{\chi}\) are denoted by \(\sigma_i\).}
    \label{fig:2-1}
\end{figure}
The form of a specific potential depends strongly upon the needs of the given application. Unfortunately,
there is currently a lack of meaningful experimental data regarding the mechanics of volumetric growth.
We therefore choose a potential that proofed to be able to predict our macroscopical observations and
further satisfies the general requirements stated above. For this purpose, the quadratic potential as
described e.g. by \cite{Stassi_1967} and \cite{Tschoegl_1971} is used in the following. This potential
can be expressed in terms of \(\omega_{hom} = m\sigma_g^2\) including the material parameters \(m\)
and \(\sigma_g\), i.e.
\begin{equation}
    \label{eq:18}
    \Phi = 3J_2 - (1-m)\sigma_g I_1 - m\sigma_g^2.
\end{equation}
As shown in Figure~\ref{fig:2-1-a}, the homeostatic state defined by this particular growth potential
forms a hyperbolic surface within the principal stress space. The tipping point of this parabola is
located on the hydrostatic axis, where its precise location is determined by the parameter \(m\)
(see Figures~\ref{fig:2-1-b} and \ref{fig:2-1-c}). From Equation~\eqref{eq:18} it is obvious that both
parameters must always be greater than zero. It is furthermore important to notice that the opening side
of the parabolic potential lies within the compressive regime for \(m< 1\) and in the tensional regime
for \(m> 1\), respectively. Since a choice of \(m=1\) describes a von-Mises type model such a choice of
this parameters must be avoided. Using this specific form of the growth potential yields the growth direction
tensor as
\begin{equation}
    \label{eq:19}
    \mathbf{N} = \mathbf{F}_g\left(3\operatorname{dev}\left(\mathbf{\Sigma}\mathbf{C}_g\right)- \left(1-m\right)\sigma_g\mathbf{I}\right)\mathbf{F}_g^{-1}.
\end{equation}
To complete the set of equations needed to describe the evolution of the growth related right Cauchy-Green
tensor, we furthermore define a particular form for the evolution of the growth multiplier \(\dot{\lambda}_g\).
From a physically motived point of view, it seems natural that the growth response increases with the
deviation of the current stress state from homeostasis. We therefore assume the change in accumulated
growth stretch is proportional to the current value of the growth potential. This furthermore ensures
that the growth process stops as soon as homeostasis is reached. With these assumptions in mind, we
choose the well established approach proposed in \cite{Perzyna_1966} and \cite{Perzyna_1971}, i.e.
\begin{equation}
    \label{eq:20}
    \dot{\lambda}_g := \frac{1}{\eta}\left(\frac{\Phi}{m\sigma_g^2}\right)^{\frac{1}{\nu}}.
\end{equation}
Herein the growth multiplier is defined in terms of the growth relaxation time \(\eta\) as well as a
non-linearity parameter \(\nu\).

\subsubsection{Choice of Helmholtz free energy}
\label{sec:2-4-3}
Until this point, the constitutive framework presented herein has been described without defining a
particular form of the Helmholtz free energy. In general, the choice of the energy potential depends
upon the specific type of material one would want to model. For the time being, we choose a compressible
Neo-Hookean type model to describe the elastic response of the material. Therefore, the elastic energy
\(\psi_e\) is written in terms of the Lam\'e constants \(\mu\) and \(\Lambda\) as
\begin{equation}
    \label{eq:21}
    \psi_e = \frac{\mu}{2}\left(\operatorname{tr}\mathbf{C}_e - 3\right) - \mu\operatorname{ln}J_e +
    \frac{\Lambda}{4}\left(J_e^2 - 1 - 2\operatorname{ln}J_e\right).
\end{equation}
Following the argumentation in Section~\ref{sec:2-2}, we furthermore define the growth related Helmholtz free
energy \(\psi_g\) in terms of a stiffness like material parameter \(\kappa_g\) such that
\begin{equation}
    \label{eq:22}
    \psi_g = \frac{\kappa_g}{2}\left(J_g^2 - 1 - 2\operatorname{ln}J_g\right).
\end{equation}
This particular choice of the growth related energy obviously fulfills the general requirements for the
definition of a strain energy density, i.e. \(\psi_g(J_g \rightarrow 0) \rightarrow \infty\) as well as
\(\psi_g(J_g = 1) = 0\) and \(\psi_g(J_g \rightarrow \infty) \rightarrow \infty\). With these definitions
at hand the second Piola-Kirchhoff stress tensor and the back-stress tensor can be derived as
\begin{equation}
    \label{eq:23}
    \begin{split}
        \mathbf{S} &= \mu\left(\mathbf{C}_g^{-1} - \mathbf{C}^{-1}\right) + \frac{\Lambda}{2}\left(\left(\frac{J}{J_g}\right)^2 - 1\right)\mathbf{C}^{-1}\\
        \mathbf{X} &= \kappa_g\left(J_g^2-1\right)\mathbf{C}_g^{-1}
    \end{split}
\end{equation}
Notice that the conjugated driving force \(\mathbf{\Gamma}\) can be easily computed, if \(\mathbf{C}_g\) and
\(\mathbf{S}\) are known (see Section~\ref{sec:2-3}).

%%% Local Variables: 
%%% mode: latex
%%% TeX-master: "../main"
%%% End: 

\section{Numerical implementation}
\label{sec:3}
To incorporate the volumetric growth model at hand into a finite element simulation framework,
a suitable time integration technique has to be used for evolution equation~\eqref{eq:16}.
As shown for example by \cite{Weber_1990}, \cite{Simo_1992}, \cite{Reese_Govindjee_1998}, \cite{Vladimirov_EtAl_2008}
and discussed in further detail by \cite{Korelc_2014}, the exponential mapping algorithm is a
very suitable choice for the treatment of the given evolution equation. We will therefore briefly
describe this approach in the following.

Starting with the introduction of discrete time increments \(\Delta t = t_{n+1} - t_n\) together with
the growth increments \(\Delta\lambda_g = \Delta t \dot{\lambda}_g\), the exponential integration
scheme for the evolution Equation~\eqref{eq:16} can be written as
\begin{equation}
    \label{eq:24}
    \mathbf{C}_{g_{n+1}} = \operatorname{exp}\left(\Delta\lambda_g\mathbf{g}\right)\mathbf{C}_{g_n}.
\end{equation}
Notice, that subscript \(n+1\) will be dropped in the following for notational simplicity, which means that any
discrete quantity without subscript will be associated with the current time step. Following the argumentation
within \cite{Vladimirov_EtAl_2008} and \cite{Dettmer_Reese_2004} Equation~\eqref{eq:24} can be reformulated to
ensure the symmetry of \(\mathbf{C}_g\). Furthermore, the authors mentioned above show that the exponential function
within this equation can be expressed in terms of the growth related right stretch tensor
\(\mathbf{U}_g = \sqrt{\mathbf{C}_g}\). Consequently, this leads to the discretized evolution equation given as
\begin{equation}
    \label{eq:25}
    \mathbf{C}_{g_n}^{-1} = \mathbf{U}_g^{-1}\operatorname{exp}\left(\Delta\lambda\mathbf{U}_g^{-1}\mathbf{f}\mathbf{U}_g^{-1}\right)\mathbf{U}_g^{-1}.
\end{equation}
In order to complete the set of discrete constitutive equations, the discrete growth multiplier \(\Delta\lambda_g\)
must determined. This can be achieved by reformulating Equation~\eqref{eq:20} (see e.g. \cite{Simo_Hughes_1998}
and \cite{deSouza_neto_2008}), i.e.
\begin{equation}
    \label{eq:26}
    \Phi = m\sigma_g^2\left(\Delta\lambda_g\eta\right)^\nu.
\end{equation}
Since both of the discrete constitutive equations are non-linear in their arguments, a local iterative solution
algorithm must be applied at integration point level to solve for both, the internal variable \(\mathbf{U}_g^{-1}\)
as well as the growth increment \(\Delta\lambda_g\). It is convenient for such algorithms to write the evolution
equations in terms of a set of coupled residual functions, which read in the case of this material model
\begin{equation}
    \label{eq:27}
    \begin{split}
        \mathbf{r}_g &= -\mathbf{C}_{g_n}^{-1} + \mathbf{U}_g^{-1}\operatorname{exp}\left(\Delta\lambda\mathbf{U}_g^{-1}\mathbf{f}\mathbf{U}_g^{-1}\right)\mathbf{U}_g^{-1} = \mathbf{0}\\
        r_\Phi &= \Phi - m\sigma_g^2\left(\Delta\lambda_g\eta\right)^\nu = 0.
    \end{split}
\end{equation}
Due to the symmetry of \(\mathbf{U}_g\), the tensor valued residual function \(\mathbf{r}_g\) can be transformed
into Voigt notation, which is computationally more efficient than solving the full tensorial equation. When applying
a Newton-Raphson procedure to solve Equations~\eqref{eq:27}, the increments of the equations arguments can be
found by solving a linearized system of equations, i.e.
\begin{equation}
    \displaystyle
    \label{eq:28}
    \begin{pmatrix}
        \frac{\partial\hat{\mathbf{r}}_g}{\partial\hat{\mathbf{U}}_g} & \frac{\partial\hat{\mathbf{r}}_g}{\partial\Delta\lambda_g} \\
        \frac{\partial r_\Phi}{\partial\hat{\mathbf{U}}_g}            & \frac{\partial r_\Phi}{\partial\Delta\lambda_g}            \\
    \end{pmatrix}
    \Delta
    \begin{pmatrix}
        \hat{\mathbf{U}}_g \\
        \Delta\lambda_g
    \end{pmatrix}
    =
    -\begin{pmatrix}
        \hat{\mathbf{r}}_{g_n} \\
        r_ {\Phi_n}
    \end{pmatrix}.
\end{equation}
During the solution process, these increments are recomputed for every iteration step in which they are used
to update the local iteration procedure. The partial derivatives used herein are not computed analytically but
rather calculated by means of an algorithmic differentiation approach. For this, the software package \textit{AceGen},
as described e.g. in \cite{Korelc_2002} and \cite{Korelc_2009}, is being used to automatically generate source code
for the computation of the tangent operators.

Since the local material response is implicitly included within the global material tangent operator of a finite
element simulation, we furthermore need to derive this tangent in a consistent manner. Otherwise, quadratic convergence
of the global iteration scheme would not be reached. For this, one should bear in mind that the second Piola-Kirchhoff
stress is a function of the right Cauchy-Green tensor as well as the internal variables. Within the given framework,
the material tangent operator can be expressed as
\begin{equation}
    \label{eq:29}
    \mathbb{C} = 2\left(
    \left.\frac{\partial\mathbf{S}}{\partial\mathbf{C}}\right\vert_{\mathbf{U}_g} +
    \left.\frac{\partial\mathbf{S}}{\partial\mathbf{U}_g}\right\vert_{\mathbf{C}} :
    \frac{\partial\mathbf{U}_g}{\partial\mathbf{C}}
    \right).
\end{equation}
Similar to the local tangent operator, these partial derivatives are computed using the software package \textit{AceGen}.
For this, the partial derivative of the growth related stretch tensor \(\mathbf{U}_g\) with respect to the right Cauchy Green
tensor can be determined from the following relation
\begin{equation}
    \label{eq:30}
    \Delta
    \begin{pmatrix}
        \hat{\mathbf{U}}_g \\
        \Delta\lambda_g
    \end{pmatrix}
    = -
    \begin{pmatrix}
        \frac{\partial\hat{\mathbf{r}}_g}{\partial\hat{\mathbf{U}}_g} & \frac{\partial\hat{\mathbf{r}}_g}{\partial\Delta\lambda_g} \\
        \frac{\partial r_\Phi}{\partial\hat{\mathbf{U}}_g}            & \frac{\partial r_\Phi}{\partial\Delta\lambda_g}            \\
    \end{pmatrix}^{-1}
    \begin{pmatrix}
        \frac{\partial\hat{\mathbf{r}}_g}{\partial\hat{\mathbf{C}}} \\
        \frac{\partial r_\Phi}{\partial\hat{\mathbf{C}}}
    \end{pmatrix}
    \Delta\hat{\mathbf{C}}.
\end{equation}
Here, we reuse the fully converged residual and jacobian from the local solution process.
Then, the desired partial derivative is given as the corresponding \(6\times 6\) submatrix located in the
upper left corner of the right-hand side matrix product.

%%% Local Variables: 
%%% mode: latex
%%% TeX-master: "../main"
%%% End: 

\section{Numerical examples}
\label{sec:4}
In the following section, numerical examples are presented to examine and discuss various aspects of the material model
introduced above. Since there is a lack of meaningful experimental data describing the volumetric growth in bioengineered
tissues, the material parameters used for the presented studies are chosen by means of an educated guess.
First, we show the influence of boundary conditions on the development of the volumetric growth process
using a simple block model. For this purpose, volumetric growth in the absence of geometrically constraining
boundary conditions is evaluated as well as the impact of both, temporal constant and time dependent constraining boundary conditions.
Next, we investigate the influence of the introduced set of material parameters, before showing structural examples of a
shrinking tissue stripe and comparing its growth related response to an isotropic growth formulation.
For the finite element simulations, we implemented the presented material model as well as the element formulation itself
into the \textit{FEAP} software package (\cite{FEAP_manual}) in terms of a \textit{user-element} routine.
For meshing and visualization of the structural examples we have used the open source software tools \textit{GMSH}
(\cite{GMSH}) and \textit{Paraview} (\cite{PARAVIEW}). Furthermore, the open source parallelisation tool \textit{GNU Parallel}
(\cite{GNUPARALLEL}) was used during evaluations of the examples shown below.

\begin{figure}%[H]
    \centering
    \begin{tikzpicture}
	%%====================================================================================================================================================================%%
	%% Coordinates
	%%====================================================================================================================================================================%%

	%% Vanishing points for perspective handling
	\coordinate (P1) at (-16cm,1.5cm); % left vanishing point (To pick)
	\coordinate (P2) at (8cm,1.5cm); % right vanishing point (To pick)

	%% (A1) and (A2) defines the 2 central points of the cuboid
	\coordinate (A1) at (0em,0.5cm); % central top point (To pick)
	\coordinate (A2) at (0em,-4.0cm); % central bottom point (To pick)

	%% (A3) to (A8) are computed given a unique parameter (or 2) .8
	% You can vary .8 from 0 to 1 to change perspective on left side
	\coordinate (A3) at ($(P1)!.8!(A2)$); % To pick for perspective 
	\coordinate (A4) at ($(P1)!.8!(A1)$);

	% You can vary .8 from 0 to 1 to change perspective on right side
	\coordinate (A7) at ($(P2)!.7!(A2)$);
	\coordinate (A8) at ($(P2)!.7!(A1)$);

	%% Automatically compute the last 2 points with intersections
	\coordinate (A5) at
	(intersection cs: first line={(A8) -- (P1)},
	second line={(A4) -- (P2)});
	\coordinate (A6) at
	(intersection cs: first line={(A7) -- (P1)},
	second line={(A3) -- (P2)});

	%% Coordinates for loading
	\coordinate (A9)  at ([yshift=50pt] A1);
	\coordinate (A10) at ($(P1)!.8!(A9)$);
	\coordinate (A12) at ($(P2)!.7!(A9)$);
	\coordinate (A11) at
	(intersection cs: first line={(A10) -- (P2)},
	second line={(A12) -- (P1)});

	%% Coordinates for boundary conditions
	\coordinate (BCL1) at ($(P1)!0.99!(A3)$);
	\coordinate (BCL2) at ($(P1)!0.99!(A4)$);
	\coordinate (BCL3) at ($(P1)!0.99!(A5)$);
	\coordinate (BCL4) at ($(P1)!0.99!(A6)$);

	\coordinate (BCR1) at ($(P2)!0.98!(A5)$);
	\coordinate (BCR2) at ($(P2)!0.98!(A6)$);
	\coordinate (BCR3) at ($(P2)!0.98!(A7)$);
	\coordinate (BCR4) at ($(P2)!0.98!(A8)$);

	%% Coordinates for coordinate system
	\coordinate (CS0) at ([xshift=-15pt, yshift=-5pt] A3);
	\coordinate (CSY) at ($(P1)!1.06!(CS0)$);
	\coordinate (CSX) at ($(P2)!1.05!(CS0)$);
	\coordinate (CSZ) at ([yshift=25pt] CS0);

	%%====================================================================================================================================================================%%
	%% DRAWINGS
	%%====================================================================================================================================================================%%

	%% Draw boundaries 
	\fill[gray!70] ([yshift=-5pt] A2) -- ([yshift=-5pt] A3) -- ([yshift=-5pt] A6) -- ([yshift=-5pt] A7) -- cycle; % face 6
	\fill[gray!70] (BCL1) -- (BCL2) -- (BCL3) -- (BCL4) -- cycle;
	\fill[gray!70] (BCR1) -- (BCR2) -- (BCR3) -- (BCR4) -- cycle;
	\node[gray] at ([xshift=-20pt] $(BCL1)!0.5!(BCL2)$) {\small \(u_y = 0\)};
	\node[gray] at ([xshift=20pt] $(BCR3)!0.5!(BCR4)$) {\small \(u_x = 0\)};
	\node[gray] at ([xshift=20pt, yshift=-10pt] $(A2)!0.5!(A7)$) {\small \(u_z = 0\)};

	%% Draw coordinate system
	\draw[->, thick] (CS0) -- (CSX);
	\draw[->, thick] (CS0) -- (CSY);
	\draw[->, thick] (CS0) -- (CSZ);
	\node[left] at (CSX) {\small \(x\)};
	\node[below] at (CSY) {\small \(y\)};
	\node[left] at (CSZ)  {\small \(z\)};

	%% Draw back faces
	\fill[white] (A2) -- (A3) -- (A6) -- (A7) -- cycle; % face 6
	\fill[white] (A3) -- (A4) -- (A5) -- (A6) -- cycle; % face 3	
	\fill[white] (A5) -- (A6) -- (A7) -- (A8) -- cycle; % face 4

	%% Transparency of block 
	%\draw[thick,dashed, gray] (A5) -- (A6);
	%\draw[thick,dashed, gray] (A3) -- (A6);
	%\draw[thick,dashed, gray] (A7) -- (A6);

	%% Draw loading
	\fill[red!30] (A9) -- (A10) -- (A11) -- (A12) -- cycle; % face 4
	\draw[->, thick, red] (A1) -- (A9);
	\draw[->, thick, red] (A4) -- (A10);
	\draw[->, thick, red] (A5) -- (A11);
	\draw[->, thick, red] (A8) -- (A12);
	\draw[red] (A9) -- (A10);
	\draw[red] (A10) -- (A11);
	\draw[red] (A11) -- (A12);
	\draw[red] (A12) -- (A9);
	\node[red, right] at (A12) {\(u_z(t)\)};

	%% Draw front lines
	\draw[thick] (A1) -- (A2);
	\draw[thick] (A3) -- (A4);
	\draw[thick] (A7) -- (A8);
	\draw[thick] (A1) -- (A4);
	\draw[thick] (A1) -- (A8);
	\draw[thick] (A2) -- (A3);
	\draw[thick] (A2) -- (A7);
	\draw[thick] (A4) -- (A5);
	\draw[thick] (A8) -- (A5);

	%% Draw Point of evaluation
	\node[blue] at ($(A1)!0.5!(A2)$) {\textbullet};
	\node[blue, below right] at ($(A1)!0.5!(A2)$) {\small{\(\text{P}_2\)}};
	\node[blue] at (A1) {\textbullet};
	\node[blue, below right] at (A1) {\small{\(\text{P}_1\)}};
\end{tikzpicture}
    \caption{Geometrical block model with uniform side length of \(1~\text{mm}\). Uniaxial boundary conditions are given in gray and time
        dependent displacement \(u_z(t)\) is denoted in red. Evaluation points \(P_1 = (1,1,1)\) and \(P_2=(1,1,0.5)\) are given in blue.}
    \label{fig:4-1}
\end{figure}
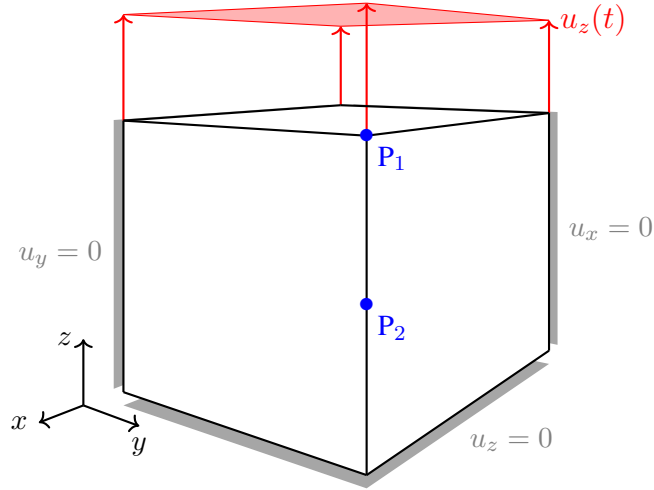

\begin{table}
    \centering
    \begin{tabularx}{\textwidth}{ r | Y Y | Y Y Y Y Y }
        \hline
                                   & \(\mu\)                                       & \(\Lambda\)                                   & \(\kappa_g\)                                  & \(m\)              & \(\sigma_g\)                                  & \(\eta\)                  & \(\nu\)            \\
                                   & \(\left[\frac{\text{N}}{\text{mm}^2}\right]\) & \(\left[\frac{\text{N}}{\text{mm}^2}\right]\) & \(\left[\frac{\text{N}}{\text{mm}^2}\right]\) & \(\left[-\right]\) & \(\left[\frac{\text{N}}{\text{mm}^2}\right]\) & \(\left[\text{s}\right]\) & \(\left[-\right]\) \\
        \hline
        Geom. unconstrained growth & 40                                            & 400                                           & 150                                           & 1.2                & 70                                            & 20                        & 1.0                \\
        Geom. constrained growth   & 40                                            & 400                                           & 250                                           & 1.2                & 200                                           & 100                       & 1.0                \\
        Clamped tissue stripe      & 100                                           & 800                                           & 150                                           & 2.0                & 250                                           & 100                       & 1.0                \\
        \hline
    \end{tabularx}
    \caption{Material parameters for numerical examples}
    \label{tbl:4-1}
\end{table}

\subsection{Geometrically unconstrained growth}
\label{sec:4-1}
\begin{figure}
    \centering
    \begin{tikzpicture}[scale=0.85]
    \begin{axis}[xmin=0, xmax=200, xlabel={Time \(t \left[\text{s}\right]\)},
            ymin=-0.4, ymax=0.0, ylabel={Stretch \(\lambda~\left[-\right]\)},
            grid=major, legend style={at={(0.5,1.25)}, anchor=north, legend columns=3}]
        \addplot[rwth1, line width = 2, mark=*, mark options={mark repeat={11}, mark size={4}}]
        table[x expr=\thisrowno{0}, y expr=\thisrowno{1}]{Simulation_data/Geom_unconstrained_growth/variation_par_m/m_1_2/Pinputa.dis};
        \addplot[rwth2, line width = 2, mark=x, mark options={mark repeat={15}, mark size={4}}]
        table[x expr=\thisrowno{0}, y expr=\thisrowno{2}]{Simulation_data/Geom_unconstrained_growth/variation_par_m/m_1_2/Pinputa.dis};
        \addplot[rwth5, line width = 2]
        table[x expr=\thisrowno{0}, y expr=\thisrowno{3}]{Simulation_data/Geom_unconstrained_growth/variation_par_m/m_1_2/Pinputa.dis};
        \legend{\(u_x\), \(u_y\), \(u_z\)}
    \end{axis}
\end{tikzpicture}
    \caption{Isotropic growth behaviour resulting in a uniform contraction in all three spatial dimensions.
        No constraining boundary conditions are applied (i.e. no \(u_z(t)\)).
        Stretches are evaluated at point \(\text{P}_1\) (see Figure~\ref{fig:4-1}).}
    \label{fig:4-2}
\end{figure}
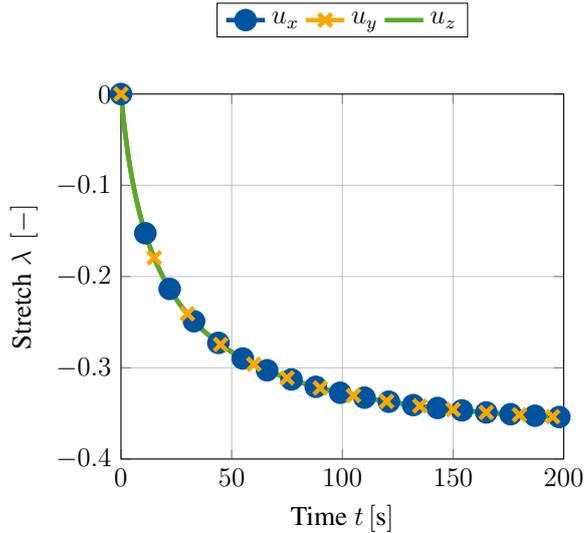
\begin{figure}
    \centering
    \begin{subfigure}{0.49\textwidth}
        \begin{tikzpicture}[scale=0.9]
    \begin{groupplot}[group style={columns=2, rows=2, horizontal sep=1.5cm, vertical sep=3cm}]
        \nextgroupplot[xmin=0,xmax=500,xlabel={Time \(t\left[\text{s}\right]\)},
            ymin=-1.0, ymax=0.0, ylabel={Stretch \(\lambda_z~\left[-\right]\)},
            grid=major, legend style={at={(0.5,1.25)}, anchor=north, legend columns=2}]
        \addplot[ rwth1, line width=2pt]
        table[x expr=\thisrowno{0}, y expr=\thisrowno{2}]{Simulation_data/Geom_unconstrained_growth/variation_par_kappa/kappa_125/Pinputa.dis};
        \addplot[ black, line width=2pt]
        table[x expr=\thisrowno{0}, y expr=\thisrowno{2}]{Simulation_data/Geom_unconstrained_growth/reference/Pinputa.dis};
        \addplot[ rwth2, line width=2pt]
        table[x expr=\thisrowno{0}, y expr=\thisrowno{2}]{Simulation_data/Geom_unconstrained_growth/variation_par_kappa/kappa_175/Pinputa.dis};
        \addplot[ rwth5, line width=2pt]
        table[x expr=\thisrowno{0}, y expr=\thisrowno{2}]{Simulation_data/Geom_unconstrained_growth/variation_par_kappa/kappa_200/Pinputa.dis};
        \legend{\(\kappa_g=125\), \(\kappa_g=150\), \(\kappa_g=175\), \(\kappa_g=200\)}
    \end{groupplot}
\end{tikzpicture}
        \caption{Various choices of the material parameter \(\kappa_g\).}
        \label{fig:4-3-a}
    \end{subfigure}
    \hfill
    \begin{subfigure}{0.49\textwidth}
        \centering
        \begin{tikzpicture}[scale=0.9]
    \begin{groupplot}[group style={columns=2, rows=2, horizontal sep=1.5cm, vertical sep=3cm}]
        \nextgroupplot[xmin=0,xmax=140, xlabel={Time \(\left[\text{s}\right]\)},
            ymin=-0.4, ymax=0.1, ylabel={Stretch \(\lambda_z \left[-\right]\)},
            grid=major, legend style={at={(0.5,1.25)}, anchor=north, legend columns=2}]
        \addplot[ rwth1, line width=2pt]
        table[x expr=\thisrowno{0}, y expr=\thisrowno{2}]{Simulation_data/Geom_unconstrained_growth/variation_par_m/m_0_5/Pinputa.dis};
        \addplot[ rwth2, line width=2pt]
        table[x expr=\thisrowno{0}, y expr=\thisrowno{2}]{Simulation_data/Geom_unconstrained_growth/variation_par_m/m_0_8/Pinputa.dis};
        \addplot[ black, line width=2pt]
        table[x expr=\thisrowno{0}, y expr=\thisrowno{2}]{Simulation_data/Geom_unconstrained_growth/reference/Pinputa.dis};
        \addplot[ rwth5, line width=2pt]
        table[x expr=\thisrowno{0}, y expr=\thisrowno{2}]{Simulation_data/Geom_unconstrained_growth/variation_par_m/m_1_5/Pinputa.dis};
        \legend{\(m=0.5\), \(m=0.8\), \(m=1.2\), \(m=1.5\)}
    \end{groupplot}
\end{tikzpicture}
        \caption{Various choices of the material parameter \(m\).}
        \label{fig:4-3-b}
    \end{subfigure}
    \hfill
    \begin{subfigure}{0.49\textwidth}
        \centering
        \begin{tikzpicture}[scale=0.9]
    \begin{groupplot}[group style={columns=2, rows=2, horizontal sep=1.5cm, vertical sep=3cm}]
        \nextgroupplot[xmin=0,xmax=500, xlabel={Time \(\left[\text{s}\right]\)},
            ymin=-1, ymax=0.0, ylabel={Stretch \(\lambda_z \left[-\right]\)},
            grid=major, legend style={at={(0.5,1.25)}, anchor=north, legend columns=2}]
        \addplot[ rwth1, line width=2pt]
        table[x expr=\thisrowno{0}, y expr=\thisrowno{2}]{Simulation_data/Geom_unconstrained_growth/variation_par_sig/sig_50/Pinputa.dis};
        \addplot[ black, line width=2pt]
        table[x expr=\thisrowno{0}, y expr=\thisrowno{2}]{Simulation_data/Geom_unconstrained_growth/reference/Pinputa.dis};
        \addplot[ rwth5, line width=2pt]
        table[x expr=\thisrowno{0}, y expr=\thisrowno{2}]{Simulation_data/Geom_unconstrained_growth/variation_par_sig/sig_75/Pinputa.dis};
        \addplot[ rwth2, line width=2pt]
        table[x expr=\thisrowno{0}, y expr=\thisrowno{2}]{Simulation_data/Geom_unconstrained_growth/variation_par_sig/sig_85/Pinputa.dis};
        \legend{\(\sigma_g=50\), \(\sigma_g=70\), \(\sigma_g=75\),\(\sigma_g=85\)}
    \end{groupplot}
\end{tikzpicture}
        \caption{Various choices of the material parameter \(\sigma_g\).}
        \label{fig:4-3-c}
    \end{subfigure}
    \hfill
    \begin{subfigure}{0.49\textwidth}
        \centering
        \begin{tikzpicture}[scale=0.9]
    \begin{groupplot}[group style={columns=2, rows=2, horizontal sep=1.5cm, vertical sep=3cm}]
        \nextgroupplot[xmin=0,xmax=500, xlabel={Time \(\left[\text{s}\right]\)},
            ymin=-0.4, ymax=0.0, ylabel={Stretch \(\lambda_z \left[-\right]\)},
            grid=major, legend style={at={(0.5,1.25)}, anchor=north, legend columns=2}]
        \addplot[ rwth1, line width=2pt]
        table[x expr=\thisrowno{0}, y expr=\thisrowno{2}]{Simulation_data/Geom_unconstrained_growth/variation_par_eta/eta_10_0/Pinputa.dis};
        \addplot[ black, line width=2pt]
        table[x expr=\thisrowno{0}, y expr=\thisrowno{2}]{Simulation_data/Geom_unconstrained_growth/reference/Pinputa.dis};
        \addplot[ rwth2, line width=2pt]
        table[x expr=\thisrowno{0}, y expr=\thisrowno{2}]{Simulation_data/Geom_unconstrained_growth/variation_par_eta/eta_50_0/Pinputa.dis};
        \addplot[ rwth5, line width=2pt]
        table[x expr=\thisrowno{0}, y expr=\thisrowno{2}]{Simulation_data/Geom_unconstrained_growth/variation_par_eta/eta_200_0/Pinputa.dis};
        \legend{\(\eta=\phantom{0}10\), \(\eta=\phantom{0}20\), \(\eta=\phantom{0}50\),\(\eta=200\)}
    \end{groupplot}
\end{tikzpicture}
        \caption{Various choices of the material parameter \(\eta\).}
        \label{fig:4-3-d}
    \end{subfigure}
    \caption{Growth induced stretch due to contraction of a block specimen for various sets of
        material parameters. No constraining boundary conditions are applied (i.e. no \(u_z(t)\)).
        Stretches are evaluated at point \(\text{P}_1\) (see Figure~\ref{fig:4-1}).}
    \label{fig:4-3}
\end{figure}
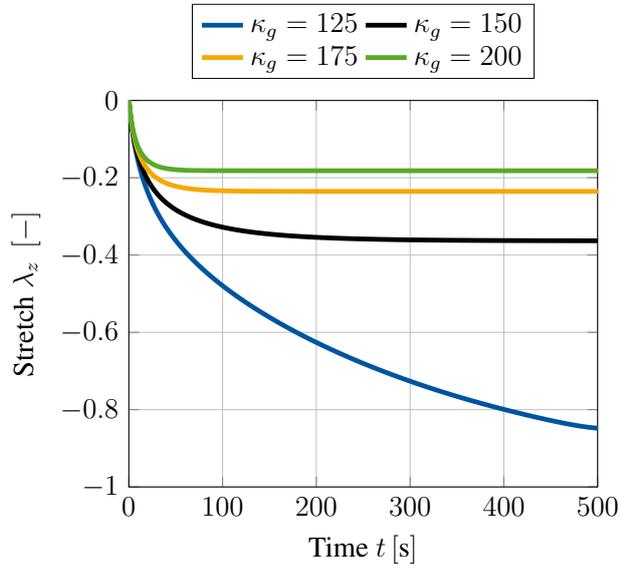
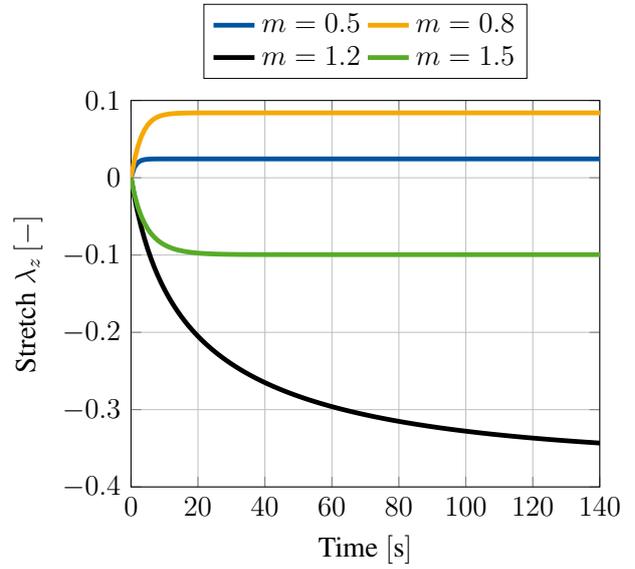
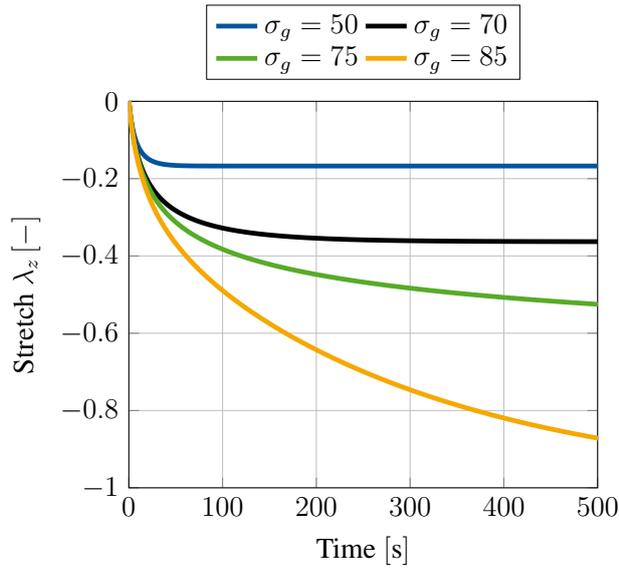
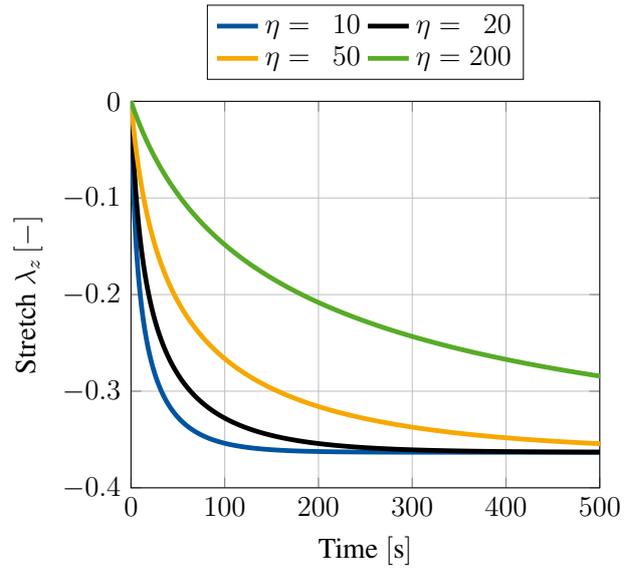

As a first example, we use the geometrical model shown in figure~\ref{fig:4-1} without applying any time dependent displacement boundary
condition \(u_z(t)\). Therefore, the specimen is able to expand or contract freely throughout the whole simulation, which should
result in an isotropic growth response. We furthermore use the set of material parameters given in Table~\ref{tbl:4-1}.
The growth response for a choice of \(m=1.2\) is visualized in
Figure~\ref{fig:4-2}. Shown by the stretches of point \(\text{P}_1\) located in the upper corner of the given block geometry,
it is obvious that the specimen contracts as expected. Since no constraining boundary conditions are applied, the overall stress
within this system should always be equal to zero and therefore could never reach a state of tensional homeostasis. It is the additional
growth related free energy, which leads to the limitation of the otherwise infinite shrinking process. One can observe this influence
really well in Figure~\ref{fig:4-3-a}, where lower values of \(\kappa_g\) lead to a more pronounced shrinking of the specimen.
It is worth noticing that \(\kappa_g > 0\) must hold for any simulation, since neglecting the contribution
of internal pressures would lead to non physical behaviour and consequently to an unstable simulation.
It is furthermore shown in Figure~\ref{fig:4-3-d} that the growth rate parameter \(\eta\) has only an impact on the speed at which
the volumetric growth process approaches the desired homeostatic state but not on its magnitude. However, as shown in Figures~\ref{fig:4-3-b}
and \ref{fig:4-3-c} a change in magnitude of the homeostatic state can be achieved by variation of \(m\) and \(\sigma_g\).
As already described in Section~\ref{sec:2-4-2}, the material parameter \(m\) defines the location of the growth potential's tipping point
on the hydrostatic axis. For values of \(m<1\) this point lies in the compressive regime, whilst a choice of \(m>1\)
pushes this point into the tension regime. As a result, the specimen approaches homeostasis either by expansion or by shrinkage.
This behaviour is really well reflected within Figure~\ref{fig:4-3-b}.
It is furthermore important to point out that for a choice of \(m=1\) the homeostatic potential
introduced in Equation~\eqref{eq:18} becomes a \textit{von Mises} type criterion, which must not be applied due to its purely
deviatoric nature. Therefore, this particular choice of \(m\) should be avoided when using the potential introduced above.

\subsection{Geometrically constrained growth}
\label{sec:4-2}
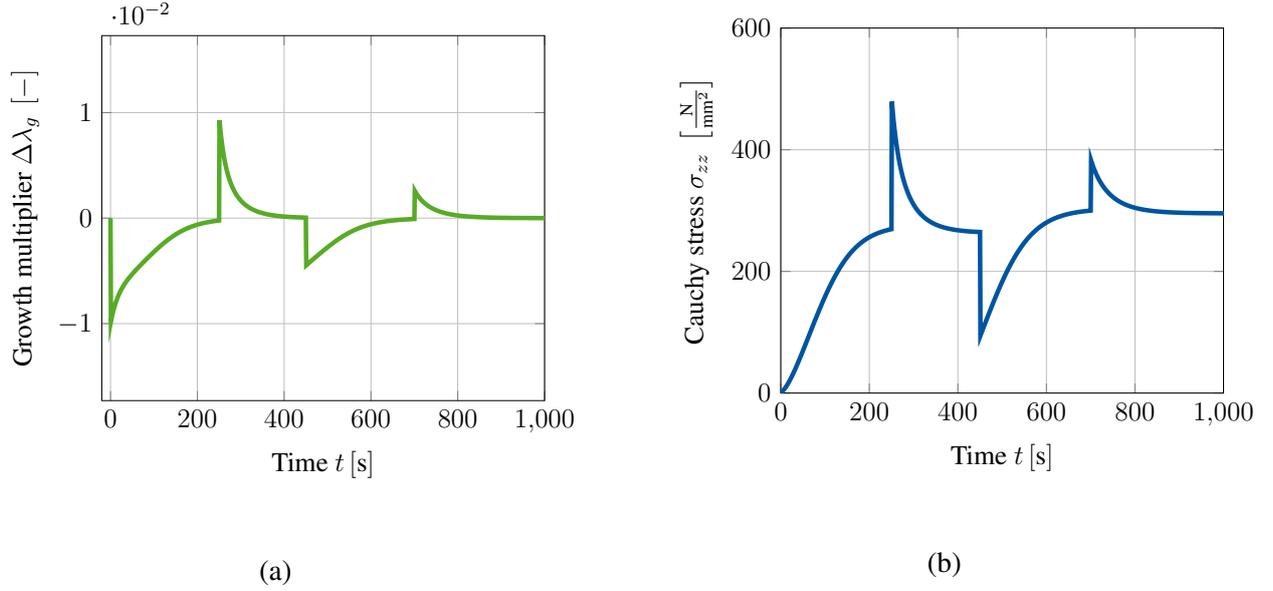
\begin{figure}
    \centering
    \begin{subfigure}{0.45\textwidth}
        \centering
        \begin{tikzpicture}[scale=0.85]
    % \begin{axis}[xmin=-20, xmax=1000, xlabel={Time \(t \left[\text{s}\right]\)},
    %         ymin=-0.35, ymax=0.35, ylabel={Stretch \(\lambda_z\left[-\right]\)},
    %         grid=major]
    %     \addplot[black, dashed, line width = 2]
    %     table[x expr=\thisrowno{0}, y expr=\thisrowno{3}]{Simulation_data/Geom_constrained_growth/Simple_tension/Pinputa.dis};
    % \end{axis}

    \begin{axis}[xmin=-20, xmax=1000, xlabel={Time \(t \left[\text{s}\right]\)},
            ymin=-0.0173, ymax=0.0173, ylabel={Growth multiplier \(\Delta\lambda_g~\left[-\right]\)}, grid=major]
        \addplot[rwth5, line width = 2]
        table[x expr=\thisrowno{0}, y expr=\thisrowno{1}]{Simulation_data/Geom_constrained_growth/Simple_tension/Pinputa.str};
    \end{axis}
\end{tikzpicture}
        \caption{}
        \label{fig:4-5-a}
    \end{subfigure}
    \hfill
    \begin{subfigure}{0.45\textwidth}
        \centering
        \begin{tikzpicture}[scale=0.85]

    \begin{axis}[xmin=0, xmax=1000, xlabel={Time \(t \left[\text{s}\right]\)},
            ymin=0, ymax=600, ylabel={Cauchy stress \(\sigma_{zz}~\left[\frac{\text{N}}{\text{mm}^2}\right]\)},
            grid=major]
        \addplot[rwth1, line width = 2]
        table[x expr=\thisrowno{0}, y expr=\thisrowno{2}]{Simulation_data/Geom_constrained_growth/Simple_tension/Pinputa.str};
    \end{axis}

\end{tikzpicture}
        \caption{}
        \label{fig:4-5-b}
    \end{subfigure}
    \caption{Evolution of Cauchy stress \(\sigma_{zz}\) and growth multiplier \(\Delta\lambda_g\) during stepwise
        loading of block specimen with \(u_z(t)\).
        Both quantities are evaluated at point \(\text{P}_1\)(Figure~\ref{fig:4-1}).
        \textit{Right}: The stress response is always converging towards a homeostatic state. This state is
        slightly different, after coming out of the compressive regime. This can be explained by the accumulted
        internal pressures described by the energy \(\psi_g\). \textit{Left}: Growth multiplier indicating, that
        the specimen is either expanding or shrinking to reach homeostasis.
    }
    \label{fig:4-5}
\end{figure}
For the next example, we choose a stepwise time dependent displacement \(u_z(t)\) to which the block given in Figure~\ref{fig:4-1} is
subjected. For the first \(250\) time steps, the displacement is held constant at \(u_z(t) = 0~\text{mm}\) before being raised to \(u_z(t)=0.3~\text{mm}\)
and held constant for another \(200\) time steps. Next, we apply compression by setting \(u_z(t)=-0.1~\text{mm}\) and holding it constant for
another \(250\) time steps. At last, \(u_z(t)\) is reset to zero again. The material parameters for this example are given in Table~\ref{tbl:4-1}.

When applying this stepwise alternating stretch to the given block specimen, it can be seen in
Figure~\ref{fig:4-5} that the material shrinks and expands depending on the current loading state,
respectively. During the first loading period, the accumulated Cauchy stress \(\sigma_{zz}\)
rises to a value of approximately \(300~\text{MPa}\), which is due to a contraction induced by the volumetric growth process. This effect is
represented by the evolution of the growth multiplier as shown in Figure~\ref{fig:4-5-a}. Since the multiplier is negative, the specimen
approaches homeostasis by shrinking. Once the displacement is raised to \(u_z(t)=0.3~\text{mm}\), the Cauchy stress \(\sigma_{zz}\) also rises
abruptly before decaying and approaching the same homeostatic stress state as before. This kind of stress reduction is achieved by an
expansion of the specimen, which is represented by a positive value of the growth multiplier. The following compression of the specimen
causes a negative jump in the overall stress response. This again induces shrinkage of the specimen in order to regain the homeostatic state
of approximately \(300~\text{MPa}\). It is important to notice that this homeostatic state is slightly higher than the state reached in the
loading cycles before. This change is due to the accumulated internal pressures described by the growth related energy \(\psi_g\).
Consequently, this results in a shift of the homeostatic surface similar to kinematic hardening in plasticity.
To what extent this effect corresponds to experimental studies is still unclear due to the lack of available data. However, there is no question
that this effect can be adapted to any experimental data without further problems by extending the model, e.g. by a non-linear formulation.
When setting \(u_z(t) = 0~\text{mm}\) in the last loading cycle, the Cauchy stresses overshoot this new homeostatic state slightly. This again results
in an expansion of the specimen in order to release the excessive stresses.

\subsection{Growth of a clamped tissue stripe}
\label{sec:4-4}
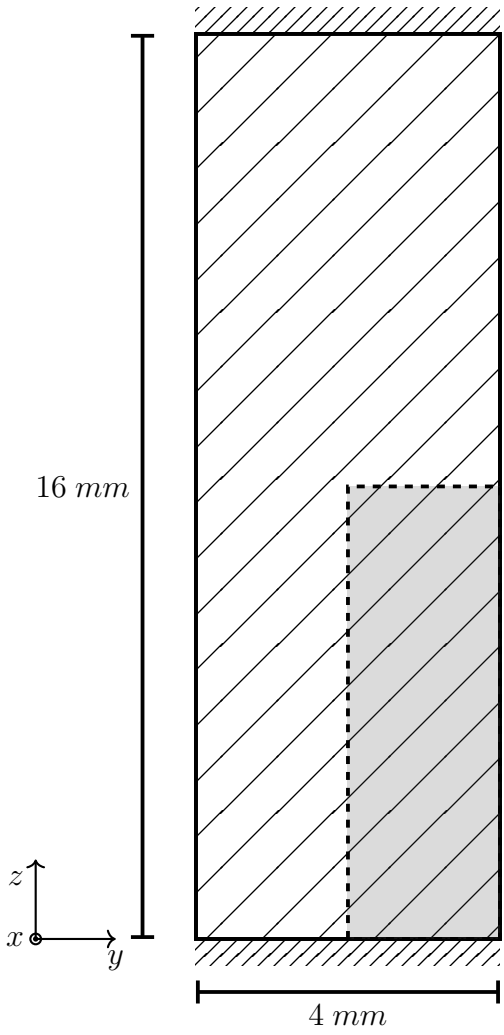
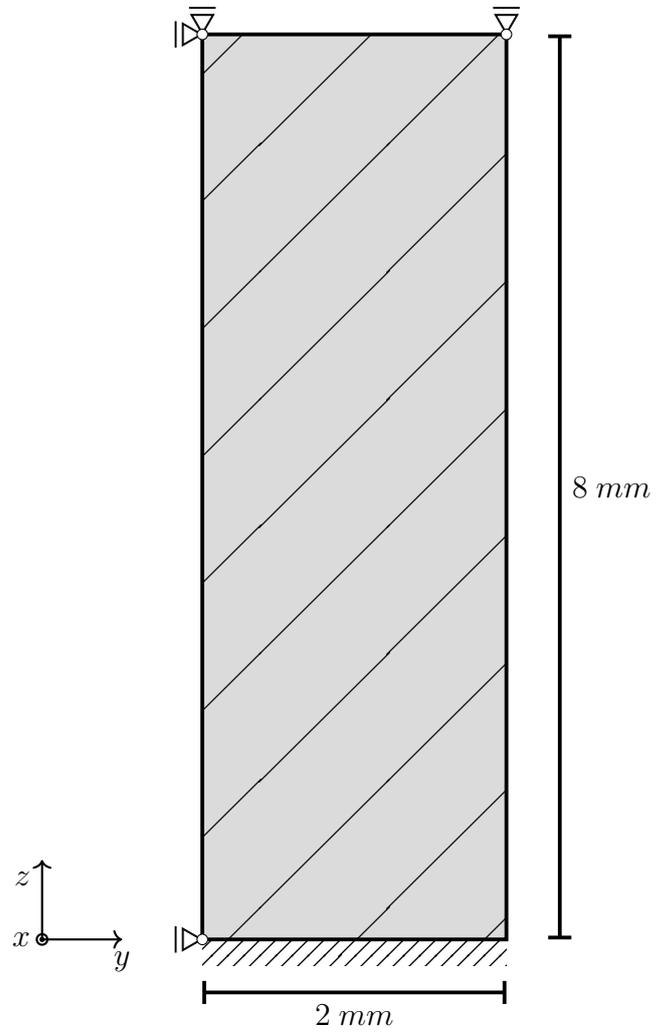
\begin{figure}
    \centering
    \begin{subfigure}{0.45\textwidth}
        \centering
        \begin{tikzpicture}
	%%====================================================================================================================================================================%%
	%% Coordinates
	%%====================================================================================================================================================================%%
	  
	%% Large rectangle
	\coordinate (P1) at (-2,  0); 
	\coordinate (P2) at (-2, 12); 
	\coordinate (P3) at ( 2, 12); 
    \coordinate (P4) at ( 2,  0);
     
	%% Small rectangle
    \coordinate (P5) at ($(P1)!0.5!(P4)$);
    \coordinate (P6) at ($(P4)!0.5!(P3)$);
    
    %% Origin
    \coordinate (O) at ([xshift=-60] P1);
    \coordinate (ytip) at ($(O)!0.5!(P1)$);
    \coordinate (ztip) at ([yshift=30pt] O);

	%%====================================================================================================================================================================%%
	%% DRAWINGS
	%%====================================================================================================================================================================%%
    % Outer rectangle
    \draw[line width=0.5mm, fill=white, pattern=northeast, hatch distance=22pt, hatch thickness = 0.5pt] (P1) -- (P2) -- (P3) -- (P4) -- cycle;
    \draw[line width=0.5mm] (P1) rectangle (P3);
    
    % Symmetric part
	\fill[black!70, opacity=0.2] (P5) rectangle (P6);
	\draw[line width=0.5mm, dashed] (P5) rectangle (P6);
	
    %% Boundary conditions
    \fill[pattern=northeast, hatch distance=7pt, hatch thickness = 0.5pt] ([yshift=-10pt] P1) rectangle (P4);
    \fill[pattern=northeast, hatch distance=7pt, hatch thickness = 0.5pt] (P2) rectangle ([yshift=10pt] P3);

    %% Measures
    \draw[line width=0.5mm,|-|] ([xshift=-20pt] P1) -- ([xshift=-20pt] P2) node[midway, left] {\(16~mm\)};
    \draw[line width=0.5mm,|-|] ([yshift=-20pt] P1) -- ([yshift=-20pt] P4) node[midway, below] {\(4~mm\)};
	    
	%% Coordinate system
    \draw[->, thick] (O) -- (ytip) node[anchor=north] {\(y\)};
    \draw[->, thick] (O) -- (ztip) node[anchor=north east] {\(z\)};
	\fill[black] (O) circle (1pt);
    \draw[thick, black] (O) circle (2pt) node[anchor=east] {\(x\)};

\end{tikzpicture}
        \caption{Full structure}
    \end{subfigure}
    \hfill
    \begin{subfigure}{0.45\textwidth}
        \centering
        \begin{tikzpicture}
	%%====================================================================================================================================================================%%
	%% Coordinates
	%%====================================================================================================================================================================%%
	  
	%% Large rectangle
	\coordinate (P1) at (-2,  0); 
	\coordinate (P2) at (-2, 12); 
	\coordinate (P3) at ( 2, 12); 
    \coordinate (P4) at ( 2,  0);
     
	%% Small rectangle
    \coordinate (P5) at ($(P1)!0.5!(P4)$);
    \coordinate (P6) at ($(P4)!0.5!(P3)$);
    
    %% Origin
    \coordinate (O) at ([xshift=-60] P1);
    \coordinate (ytip) at ($(O)!0.5!(P1)$);
    \coordinate (ztip) at ([yshift=30pt] O);

	%%====================================================================================================================================================================%%
	%% DRAWINGS
	%%====================================================================================================================================================================%%
    % Outer rectangle
    \draw[line width=0.5mm, fill=white, pattern=northeast, hatch distance=49pt, hatch thickness = 0.5pt] (P1) -- (P2) -- (P3) -- (P4) -- cycle;
    \draw[line width=0.5mm, fill=black!70, opacity=0.2] (P1) -- (P2) -- (P3) -- (P4) -- cycle;
	
    %% Boundary conditions
    \fill[pattern=northeast, hatch distance=7pt, hatch thickness = 0.5pt] ([yshift=-10pt] P1) rectangle (P4);
    
    % @ P3
    \node[mark size=5pt, rotate=-180] at ([yshift=5pt] P3) {\pgfuseplotmark{triangle*}};
    \node[mark size=3.5pt, color=white, rotate=-180,] at ([yshift=5pt] P3) {\pgfuseplotmark{triangle*}};
    \draw[fill=white] (P3) circle (0.07cm);
    \draw[line width=0.25mm] ([xshift=-5pt, yshift=10pt] P3) -- ([xshift=5pt, yshift=10pt] P3);

    % @ P2
    \node[mark size=5pt, rotate=-180] at ([yshift=5pt] P2) {\pgfuseplotmark{triangle*}};
    \node[mark size=3.5pt, color=white, rotate=-180,] at ([yshift=5pt] P2) {\pgfuseplotmark{triangle*}};
    \node[mark size=5pt, rotate=-90] at ([xshift=-5pt] P2) {\pgfuseplotmark{triangle*}};
    \node[mark size=3.5pt, color=white, rotate=-90,] at ([xshift=-5pt] P2) {\pgfuseplotmark{triangle*}};
    \draw[fill=white] (P2) circle (0.07cm);
    \draw[line width=0.25mm] ([xshift=-5pt, yshift=10pt] P2) -- ([xshift=5pt, yshift=10pt] P2);
    \draw[line width=0.25mm] ([xshift=-10pt, yshift=-5pt] P2) -- ([xshift=-10pt, yshift=5pt] P2);

    % @ P1
    \node[mark size=5pt, rotate=-90] at ([xshift=-5pt] P1) {\pgfuseplotmark{triangle*}};
    \node[mark size=3.5pt, color=white, rotate=-90,] at ([xshift=-5pt] P1) {\pgfuseplotmark{triangle*}};
    \draw[fill=white] (P1) circle (0.07cm);
    \draw[line width=0.25mm] ([xshift=-10pt, yshift=-5pt] P1) -- ([xshift=-10pt, yshift=5pt] P1);

    %% Measures
    \draw[line width=0.5mm,|-|] ([xshift=20pt] P3) -- ([xshift=20pt] P4) node[midway, right] {\(8~mm\)};
    \draw[line width=0.5mm,|-|] ([yshift=-20pt] P1) -- ([yshift=-20pt] P4) node[midway, below] {\(2~mm\)};
	    
	%% Coordinate system
    \draw[->, thick] (O) -- (ytip) node[anchor=north] {\(y\)};
    \draw[->, thick] (O) -- (ztip) node[anchor=north east] {\(z\)};
	\fill[black] (O) circle (1pt);
    \draw[thick, black] (O) circle (2pt) node[anchor=east] {\(x\)};

\end{tikzpicture}
        \caption{Symmetric part}
    \end{subfigure}
    \caption{Geometric model of clamped tissue stripe with thickness of \(t=2~\text{mm}\). The overall structure is also
        supported in the \(x\) direction.}
    \label{fig:4-6}
\end{figure}

\begin{figure}
    \centering
    \begin{subfigure}{0.19\textwidth}
        \centering
        \includegraphics[scale=0.3]{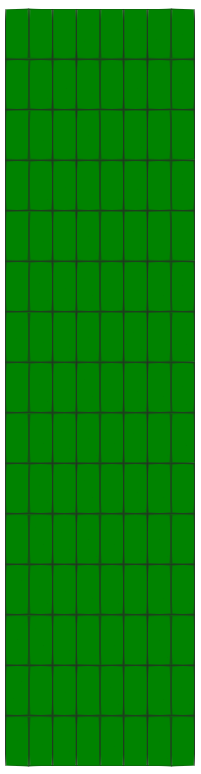}
        \caption{360 el.}
    \end{subfigure}
    \hfill
    \begin{subfigure}{0.19\textwidth}
        \centering
        \includegraphics[scale=0.3]{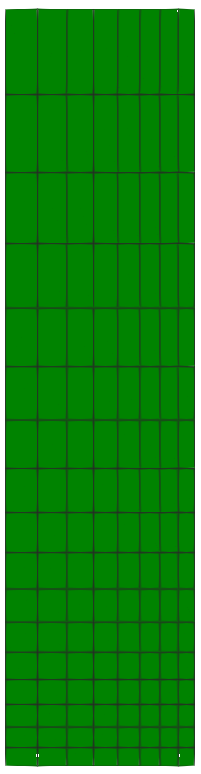}
        \caption{408 el.}
    \end{subfigure}
    \hfill
    \begin{subfigure}{0.19\textwidth}
        \centering
        \includegraphics[scale=0.3]{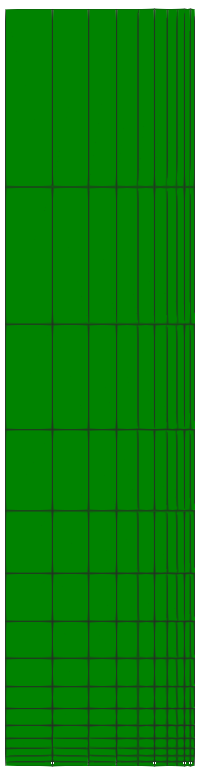}
        \caption{450 el.}
    \end{subfigure}
    \hfill
    \begin{subfigure}{0.19\textwidth}
        \centering
        \includegraphics[scale=0.3]{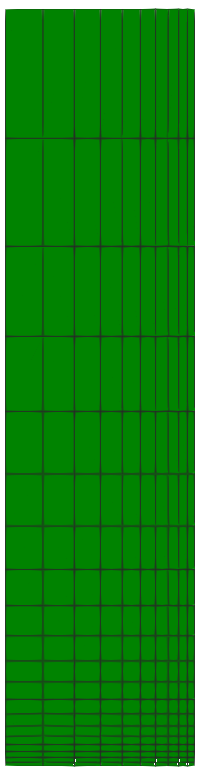}
        \caption{1000 el.}
    \end{subfigure}
    \hfill
    \begin{subfigure}{0.19\textwidth}
        \centering
        \includegraphics[scale=0.3]{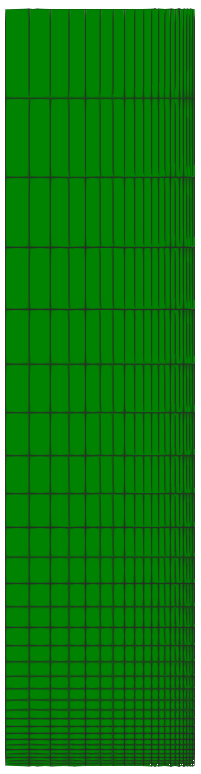}
        \caption{3000 el.}
    \end{subfigure}
    \hfill
    \begin{subfigure}{0.49\textwidth}
        \centering
        \includegraphics[scale=0.16]{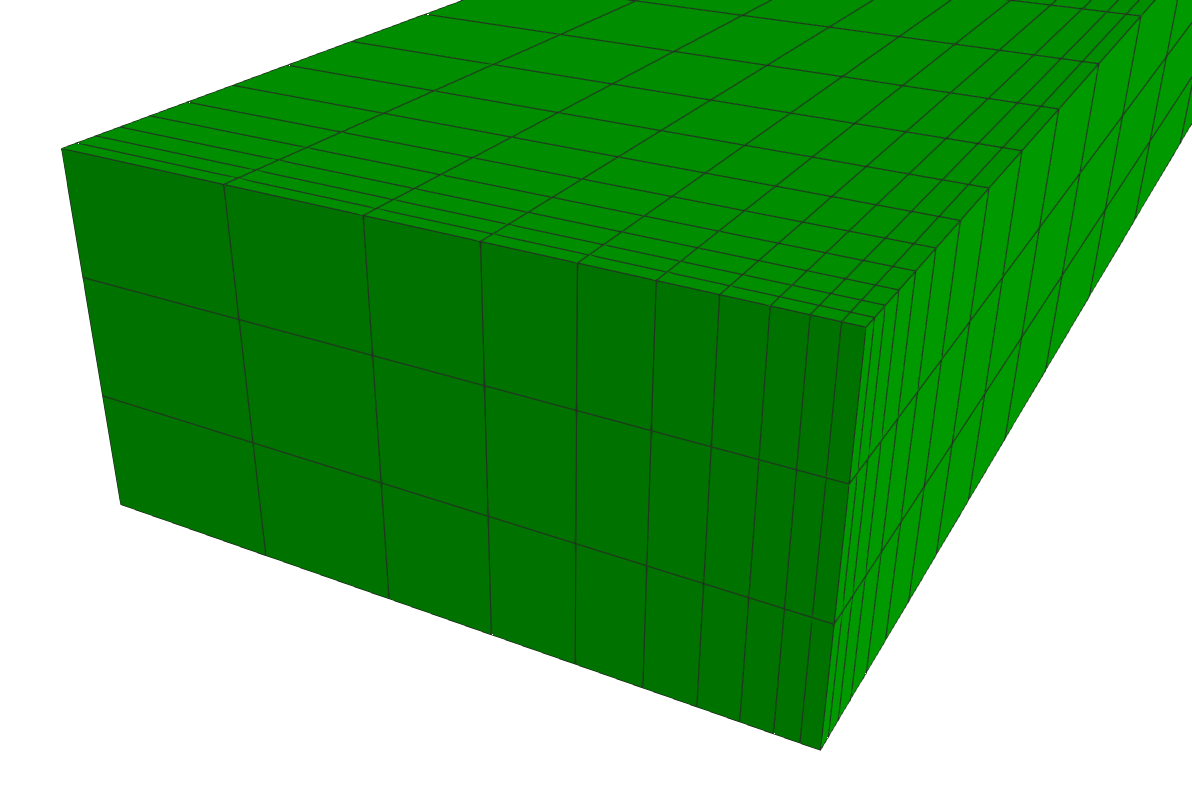}
        \caption{Close up of right corner for 450 el.}
    \end{subfigure}
    \hfill
    \begin{subfigure}{0.49\textwidth}
        \centering
        \includegraphics[scale=0.16]{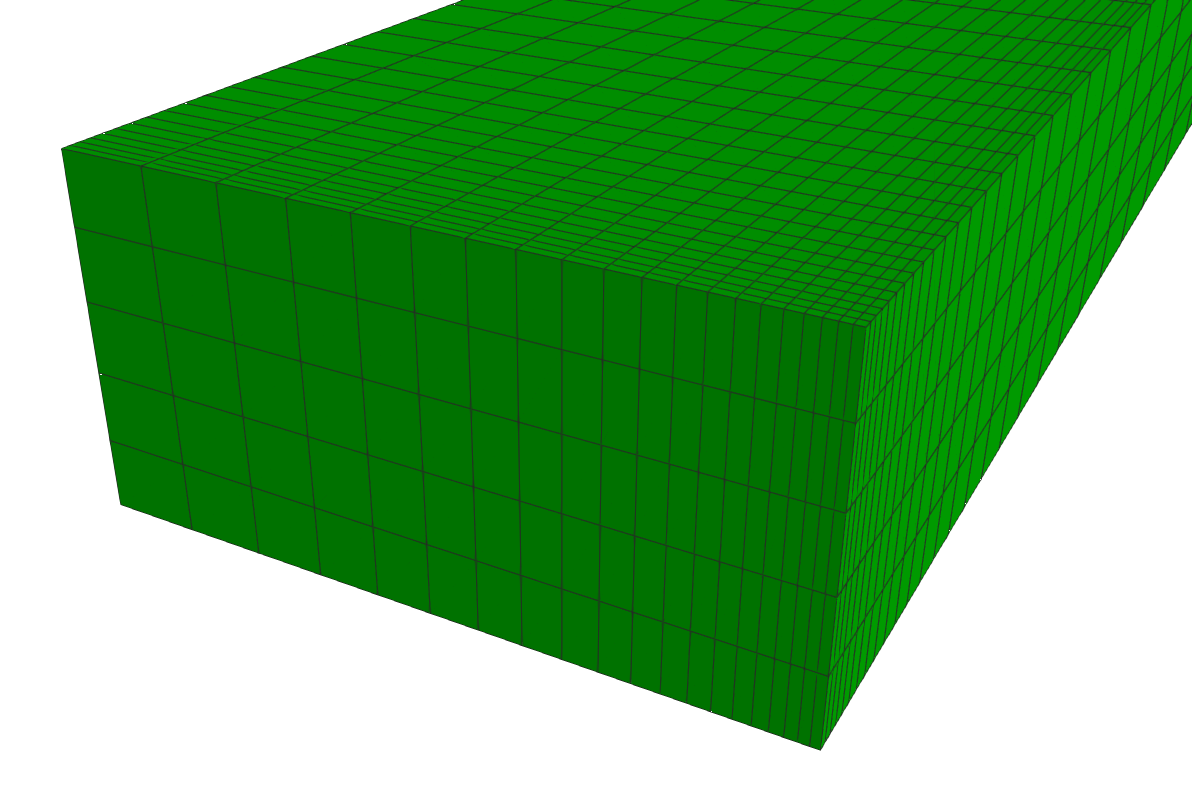}
        \caption{Close up of right corner for 3000 el.}
    \end{subfigure}
    \caption{Mesh refinements for symmetric part of clamped tissue stripe.}
    \label{fig:4-7}
\end{figure}

In the next example we consider the volumetric growth process within a tissue stripe that is clamped at both ends such that no stresses are induced
at time \(t=0\). Under these conditions, the tissue stripe is expected to shrink, which induces a homeostatic stress state that is dominated
by tension. Such effects have been shown experimentally e.g. by \cite{Ghazanfari_2015} among others. As illustrated in Figure~\ref{fig:4-6},
symmetric properties are exploited such that only a quarter of the full specimen is used for the following simulations.
The elastic and growth related material parameters applied in this example are chosen in such a way that the desired shrinkage of the specimen is achieved.
These parameters are given in Table~\ref{tbl:4-1}. For the spatial discretisation, a standard linear (Q1) finite element formulation is adopted with various meshes
containing \(360\), \(408\), \(450\), \(1000\) and \(3000\) elements (see Figure~\ref{fig:4-7}). Since the most pronounced stresses are expected to
occur in the lower right corner of the symmetric specimen, the mesh is refined with a focus on this particular region.
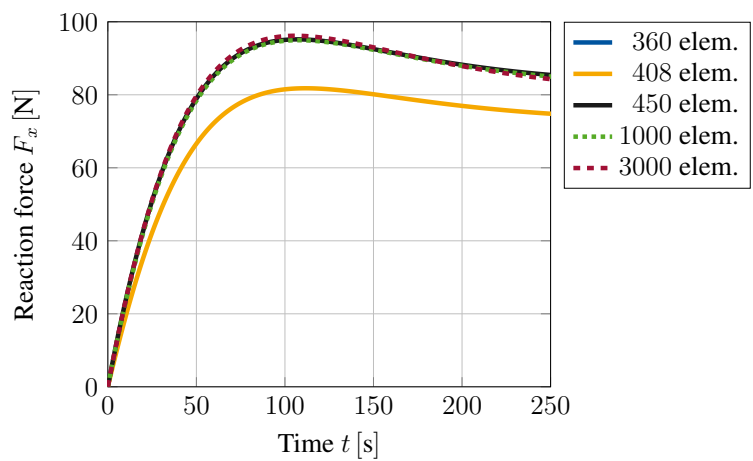
\begin{figure}
    \centering
    \begin{tikzpicture}[scale=0.85]
        \begin{groupplot}[group style={columns=1, rows=3, horizontal sep=2.5cm, vertical sep=2cm}]

                \nextgroupplot[xmin=0, xmax=250, xlabel={Time \(t \left[\text{s}\right]\)},
                        ymin=00, ymax=100, ylabel={Reaction force \(F_x\left[\text{N}\right]\)},
                        grid=major, legend pos=outer north east]
                \addplot[rwth1, line width = 2]
                table[x expr=\thisrowno{0}, y expr=\thisrowno{1}]{Simulation_data/Clamped_stripe/el_360/Pinputa.sum};
                \addplot[rwth2, line width = 2]
                table[x expr=\thisrowno{0}, y expr=\thisrowno{1}]{Simulation_data/Clamped_stripe/el_408/Pinputa.sum};
                \addplot[black!90, line width = 2]
                table[x expr=\thisrowno{0}, y expr=\thisrowno{1}]{Simulation_data/Clamped_stripe/el_450/Pinputa.sum};
                \addplot[rwth5, dotted, line width = 2]
                table[x expr=\thisrowno{0}, y expr=\thisrowno{1}]{Simulation_data/Clamped_stripe/el_1000/Pinputa.sum};
                \addplot[rwth6, dashed, line width = 2]
                table[x expr=\thisrowno{0}, y expr=\thisrowno{1}]{Simulation_data/Clamped_stripe/el_3000/Pinputa.sum};
                \legend{\(\phantom{0}360\) elem., \(\phantom{0}408\) elem., \(\phantom{0}450\) elem., \(1000\) elem., \(3000\) elem.}

        \end{groupplot}
\end{tikzpicture}
    \caption{Reaction force of a clamped tissue stripe evaluated at \(z=0\) for various mesh sizes.
        Mesh convergence can be observed nicely.}
    \label{fig:4-8}
\end{figure}
When considering the reaction force \(F_x\) evaluated over time at \(z=0\), Figure~\ref{fig:4-8} shows good convergence behaviour for increasing number
of elements within the mesh. Similar results can be obtained when evaluating the reaction forces in \(y\) and \(z\) direction, respectively. Although
the solution of a mesh containing \(450\) elements has already reached convergence, for visualization purposes, the finest discretisation containing \(3000\) elements
is used in the following.

\begin{figure}
    \centering
    \include{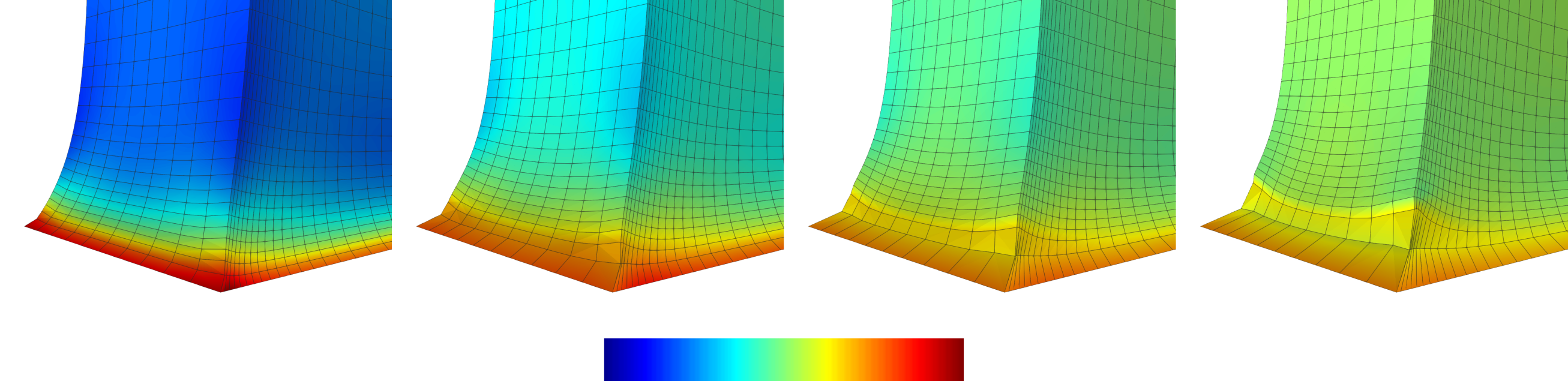}
    \caption{Pronounced distortion of elements at the clamped corner due to growth related reduction
        of exceeding stresses. Maximum principal Cauchy stresses are plotted for four different snapshots in time.}
    \label{fig:4-11}
\end{figure}

To show the capabilities of the newly introduced material model, we next compare its response to the
growth behaviour of a well established model for isotropic volumetric growth. For this, we adapted the
model of \cite{Lubara_Hoger_2002} such that it is capable of reaching a prescribed homeostatic state.
Details about the evolution equations for this particular model are given in Appendix~\ref{sec:Appendix-3}.
Within this formulation, we use the material parameter \(M_{crit}=80~\text{MPa}\) to describe the
homeostatic stress state that shall ultimately be reached. For the positive and negative growth
velocities \(k^+=0.1\) and \(k^-=0.1\) are chosen, respectively. The upper and lower growth boundaries
are set to \(\vartheta^+ = 2.0\) and \(\vartheta^-=0.25\), while the shape factors are given as
\(\gamma^+=2\) and \(\gamma^-=3\).
\begin{figure}
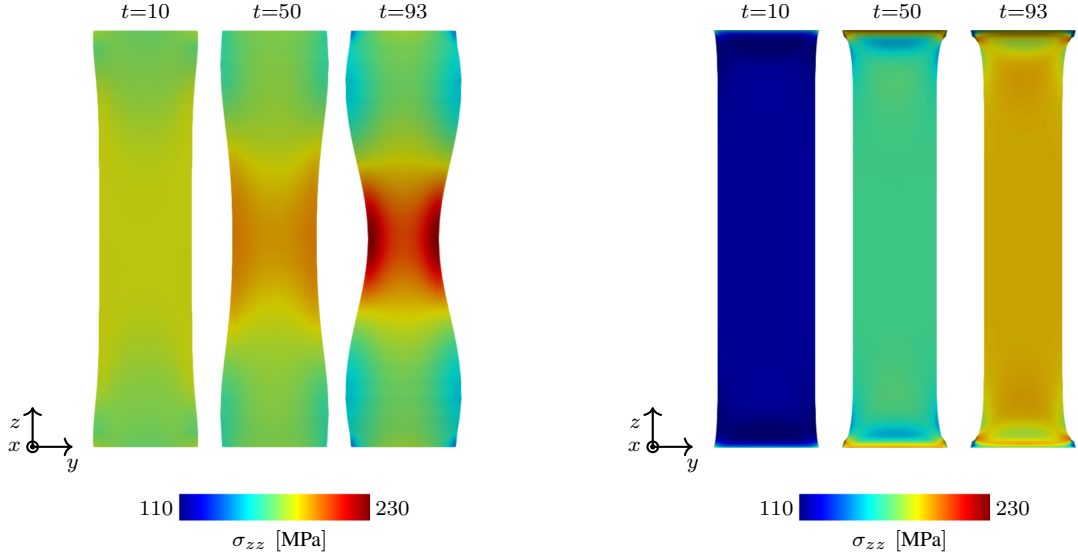

    \centering
    \begin{subfigure}{0.49\textwidth}
        \centering
        \include{images/Stripe_model/comparison_iso_aniso/isotropic}
        \caption{Isotropic growth model using a modified version of \cite{Lubara_Hoger_2002}}
        \label{fig:4-9-a}
    \end{subfigure}
    \hfill
    \begin{subfigure}{0.49\textwidth}
        \centering
        \include{images/Stripe_model/comparison_iso_aniso/anisotropic}
        \caption{New volumetric growth model using the general growth potential}
        \label{fig:4-9-b}
    \end{subfigure}
    \caption{Comparison of an isotropic growth model with the newly introduced formulation. The response of a
        clamped tissue stripe differs significantly in both, shape as well as the displayed stress response
        (Cauchy stresses \(\sigma_{zz}\)).}
    \label{fig:4-9}
\end{figure}
First of all it is important to notice that the given isotropic formulation shows severe stability problems
for the example at hand. More precisely, as soon as material parameters are chosen such that a similar
homeostatic stress state shall be reached within the specimen, the simulation becomes unstable after a
finite number of time steps and eventually breaks. When taking a closer look at the evolution of the
growth process as it is shown for three distinct time steps in Figure~\ref{fig:4-9}, it is obvious that
the starting point of the instability can be located at the clamping of the tissue stripe. Due to the
initial contraction of the overall tissue stripe, a multi-axial stress state is induced at the clamping.
In this region, the stress state soon exceeds the desired homeostatic state which yields an expansion
of the material in order to release excessive stresses. Whilst the newly derived growth model reduces this
stress state by expanding anisotropically,
the isotropic formulation seems not to be able to deal with this effect. This is due to the fact that
an isotropic growth formulation can only predict expansion or shrinkage uniformly in all three spatial
dimensions. Such a uniform expansion at the foot of the specimen results in a passive compression of the
specimens middle part, reducing the overall stress within this region and therefore inducing further
contraction. This again triggers an increasing expansion in the foot of the specimen. A vicious cycle is
born, which eventually leads to the hourglass like shape of the specimen as it is shown in
Figure~\ref{fig:4-9-a}. Ultimately, this leads to instabilities and a failing simulation at \(t = 93\).
For sure, it is possible to reduce such unwanted behaviour by variation of the material parameters.
Nevertheless, the general problem of a non-physical expansion in the foot area could not be cured with
such an approach. This example shows clearly how restrictive and, therefore, unsuitable the assumption of
isotropic growth is, even for a relatively simple structure as the one shown in this example.
Taking a closer look at the stress response of the newly derived model, one can observe the exceeding
maximum principal Cauchy stresses \(\sigma_{max}\) located at the clamped foot of the specimen
(see Figure~\ref{fig:4-11}) being released due to the anisotropic expansion process. This effect can
also be observed in Figure~\ref{fig:4-8}, where the reaction forces reach a maximum at time \(t=100\) and decrease
afterwards to approach a converged state. Unfortunately, this effect also leads to a pronounced
distortion of the associated elements within the corners of the clamped stripe. Figure~\ref{fig:4-11}
shows that this artefact is even noticable for the finest mesh evaluated. Nevertheless, it is
important to emphasize that this effect so far does not have an influence on the stability of
the given simulation. Due to the incompressible nature of the material, it is possible that
such behaviour is also amplified by shear or volumetric locking effects and would not occur in
such a pronounced manner if locking would not play a role. However, the influence of possible
locking effects is out of scope for this work.

%%% Local Variables: 
%%% mode: latex
%%% TeX-master: "../main"
%%% End: 
\section{Conclusion and outlook}
\label{sec:5}
In this paper, we developed a novel model for the description of stress driven volumetric growth. This approach is based on the well
established multiplicative split of the deformation gradient into an elastic and a growth related part. Furthermore, we made the assumption
that the given material adapts to its surroundings such that a certain homeostatic stress state is induced within the material. For this homeostatic state,
we assume that it can be described in terms of a scalar valued stress like quantity, which led us to the definition of a growth potential.
With this idea in hand, we defined an evolution law for the growth related right Cauchy-Green tensor by means of a time dependent associative
rule. This approach is similar but not identical to those often used in the field of finite visco-plasticity. With these basic modelling assumptions, we
were able to show that  this approach is capable of simulating both, isotropic and anisotropic growth behaviour within one singular formulation.
The distinction between isotropic and anisotropic response is merely a question of the applied boundary conditions and not a-priori prescribed
by the structure of the growth tensor. The advantages of this approach have been shown by comparing it to a standard formulation of isotropic growth.
In the authors' opinion, the results of the evaluations shown within this publication are very promising. Unfortunately, we were not able to validate
the given model due to the lack of experimental data published in the literature. Conducting meaningful experiments would therefore be a first step
towards validating the model described above. Since the overall framework of the model is quiet general, it seems possible to easily adapt the growth
behaviour to fit various experiments. For this, the choice of alternative descriptions for the growth potential as well as the evolution equation for
the growth multiplier could be investigated. To this point, our formulation makes use of a purely isotropic elastic ground model, i.e. Neo-Hooke.
Since biological tissue by its very own nature is composed of various components, such as e.g. collagen and elastin, the assumption of material isotropy
is not ideal. Therefore, we suggest that the given elastic ground model could be extended to also capture the anisotropic nature of the underlying
material response properly. This could be achieved by introducing an additional dependency within the Helmholtz free energy that is defined
by means of structural tensors describing e.g. the direction of collagen fibres. Furthermore, the investigation of locking effects triggered by the
nearly incompressible material behaviour of biological tissues might also be of interest. Since standard low order finite element formulations are
particularly vulnerable in this area, the finite element implementation should therefore be considered more closely. Investigating the influence of reduced
integration finite elements seems to be of high benefit. Especially the element formulations Q1SP (see \cite{Reese_2005}) or Q1STx (see \cite{Schwarze_2011},
\cite{Barfusz_2021}) could improve the computation in terms of computational accuracy as well as computational speed.

%%% Local Variables: 
%%% mode: latex
%%% TeX-master: "../main"
%%% End: 

\appendix
\section{Appendix}
\label{sec:Appendix}

\subsection{Invariants of Mandel stress tensor and Kirchhoff stress tensor}
\label{sec:Appendix-1}
The Kirchhoff stress tensor is defined as
\begin{equation*}
    \boldsymbol{\tau} = \mathbf{F}\mathbf{S}\mathbf{F}^T.
\end{equation*}
Making use of the identity \(\mathbf{C}_e\mathbf{F}_g = \mathbf{F}_g^{-T}\mathbf{C}\) and using a push forward operation
on the second Piola-Kirchhoff stress tensor \(\mathbf{S}\), the definition of the Mandel stress tensor can be rewritten as
\begin{equation*}
    \begin{split}
        \mathbf{M} &= \mathbf{C}_e \mathbf{F}_g\mathbf{S}\mathbf{F}_g^T\\
        &= \mathbf{F}_g^{-T}\mathbf{C}\mathbf{S}\mathbf{F}_g^T.
    \end{split}
\end{equation*}
With this at hand, it is easy to show that the main invariants \(J_\alpha\) with \(\alpha\in{1,2,3}\) are identical for
both, the Mandel and the Kirchhoff stress tensor, i.e.
\begin{equation*}
    \begin{split}
        J_\alpha &= \operatorname{tr}\left(\mathbf{M}^\alpha\right)\\
        &= \operatorname{tr}\left(\left(\mathbf{F}_g^{-T}\mathbf{C}\mathbf{S}\mathbf{F}_g^T\right)^\alpha\right)\\
        &= \operatorname{tr}\left(\left(\mathbf{C}\mathbf{S}\right)^\alpha\right)\\
        &= \operatorname{tr}\left(\left(\mathbf{F}\mathbf{S}\mathbf{F}^T\right)^\alpha\right)\\
        &= \operatorname{tr}\left(\boldsymbol{\tau}^\alpha\right).
    \end{split}
\end{equation*}

\subsection{Isotropic growth model for comparison}
\label{sec:Appendix-3}
The isotropic growth model used for comparison with the anisotropic growth model developed herein is based on the formulation of \cite{Lubara_Hoger_2002}.
It uses the multiplicative split of the deformation gradient, i.e.
\begin{equation*}
    \mathbf{F} = \mathbf{F}_e \mathbf{F}_g,
\end{equation*}
where the growth related deformation gradient is defined as
\begin{equation*}
    \mathbf{F}_g = \vartheta \mathbf{I},
\end{equation*}
with \(\vartheta\) describing the growth induced stretch. The evolution equation of this particular model is defined in terms of the Mandel
stress tensor \(\mathbf{M}\) as well as a set of material parameters, i.e.
\begin{equation*}
    \dot{\vartheta} := k(\vartheta)\phi(\mathbf{M}).
\end{equation*}
Here the driving force \(\phi\) is defined as
\begin{equation*}
    \phi := \operatorname{tr}\mathbf{M} - M_{crit},
\end{equation*}
where \(M_{crit}\) describes the desired homeostatic stress state that should be reached by the material. Furthermore, the growth velocity is described by
\begin{equation*}
    k(\vartheta) :=
    \begin{cases}
        k^+\left(\frac{\vartheta^+ - \vartheta}{\vartheta^+ - 1}\right)^{\gamma^+} & \text{if } \phi > 0   \\
        k^-\left(\frac{\vartheta - \vartheta^-}{1 - \vartheta^-}\right)^{\gamma^-} & \text{if } \phi < 0 ,
    \end{cases}
\end{equation*}
with \(k^+\), \(k^-\) denote the expansion and contraction speed, respectively. To restrict the growth process, the parameters \(\vartheta^+\) and \(\vartheta^-\)
are introduced as upper and lower thresholds of the growth induced stretch. Finally, two shape factors for the evolution are described by \(\gamma^+\) and \(\gamma^-\).
For further information on this particular model, the reader is kindly referred to the original publication.

\subsection{Transformation of invariants from intermediate to reference configuration}
\label{sec:Appendix-2}
Reformulating the definition of the referential driving force, i.e.
\begin{equation*}
    \mathbf{M} - \boldsymbol{\chi} = \mathbf{F}_g\mathbf{\Sigma}\mathbf{F}_g^{T},
\end{equation*}
directly yields the new definition for the volumetric invariant, i.e.
\begin{equation*}
    \begin{split}
        I_1 &= \operatorname{tr}\left(\mathbf{M} - \boldsymbol{\chi}\right)\\
        &= \operatorname{tr}\left(\mathbf{F}_g\mathbf{\Sigma}\mathbf{F}_g^{T}\right)\\
        & = \operatorname{tr}\left(\mathbf{\Sigma}\mathbf{C}_g\right).
    \end{split}
\end{equation*}
Making use of this relation, the deviatoric part of the driving force is given by
\begin{equation*}
    \begin{split}
        \operatorname{dev}\left(\mathbf{M} - \boldsymbol{\chi}\right) &= \operatorname{dev}\left(\mathbf{F}_g\mathbf{\Sigma}\mathbf{F}_g^{T}\right)\\
        &= \mathbf{F}_g\mathbf{\Sigma}\mathbf{F}_g^{T} - \frac{1}{3}\operatorname{tr}\left(\mathbf{F}_g\mathbf{\Sigma}\mathbf{F}_g^{T}\right)\mathbf{I}\\
        &= \mathbf{F}_g\mathbf{\Sigma}\mathbf{C}_g\mathbf{F}_g^{-1} - \frac{1}{3}\operatorname{tr}\left(\mathbf{\Sigma}\mathbf{C}_g\right)\mathbf{I}\\
        &= \mathbf{F}_g\left(\mathbf{\Sigma}\mathbf{C}_g - \frac{1}{3}\operatorname{tr}\left(\mathbf{\Sigma}\mathbf{C}_g\right)\mathbf{I}\right)\mathbf{F}_g^{-1}\\
        &= \mathbf{F}_g\operatorname{dev}\left(\mathbf{\Sigma}\mathbf{C}_g\right)\mathbf{F}_g^{-1}.
    \end{split}
\end{equation*}
Utilizing the properties of the trace operator, the deviatoric invariant can be rewritten as
\begin{equation*}
    J_2 = \frac{1}{2}\left(\operatorname{dev}\left(\mathbf{\Sigma}\mathbf{C}_g\right)^2\right).
\end{equation*}

\section{Declarations}
\subsection{Funding}
This work was funded through the grants RE 1057/45-1 (No. 403471716) and RE 1057/46-1 (No. 404502442)
of the German Research Foundation (DFG). Furthermore the AiF
grant provided under the project number IGF 21348 N/3 was part of funding this work.

\subsection{Conflict of interest}
The authors of this work certify that they have no affiliations with or involvement in any
organization or entity with any financial interest (such as honoraria; educational grants;
participation in speakers’ bureaus; membership, employment, consultancies, stock ownership,
or other equity interest; and expert testimony or patent-licensing arrangements), or non-financial
interest (such as personal or professional relationships, affiliations, knowledge or beliefs) in
the subject matter or materials discussed in this manuscript.

\subsection{Availability of data and material}
The generated data is stored redundantly and will be made available on demand.

\subsection{Code availability}
The custom routines will be made available on demand. The software package FEAP is a proprietary
software and can therefore not be made available.

\subsection{Author's contributions}
L. Lamm reviewed the relevant existing literature, performed all simulations, interpreted the results and
wrote this article. L. Lamm and H. Holthusen worked out the theoretical material model and impemented it
into the finite element software FEAP. H. Holthusen, T. Brepols, S. Jockenh\"ovel and S. Reese gave conceptual
advice, contributed in the discussion of the results, read the articel and gave valuable suggestions for
improvement. All authors approved the publication of the manuscript.

%%%%%%%%%%%%%%%%%%%%%%%%%%%%%%%%%%%%%%%%%%%%%%%%%%%%%%%%%%%%%%%%%%%%%%%%%%%%%%%%%%%%%%%%%%%%%%%%%%%%%%%%%

\bibliographystyle{agsm}
\bibliography{literature}

\end{document}